\documentclass[reprint,amsmath,amssymb,aps,prl,noeprint,twoside]{revtex4-2}

\def\Publication{1}

\usepackage{graphicx}
\usepackage{dcolumn}
\usepackage[caption=false]{subfig}

\usepackage{bm}
\newcommand{\tnn}{\ifdefined\tJConvention t_1\else t\fi}
\newcommand{\tnnn}{\ifdefined\tJConvention t_2\else t'\fi}
\newcommand{\tnnns}{\ifdefined\tJConvention t_2\else t'\!\fi}
\newcommand{\eA}{\ifdefined\Publication e_1\else e_{\!\!\;A}\fi}
\newcommand{\eS}{\ifdefined\Publication e_2\else e_S\fi}

\newcommand{\ey}{\ifdefined\Publication \mathbf{y}\else y\fi}

\usepackage{setspace}
\usepackage{hyperref}
\hypersetup{colorlinks=true,
linkcolor=blue,
filecolor=blue,      
urlcolor=blue,
citecolor=blue,
pdftitle={Frustration Induced Superconductivity in the Hubbard Model},
pdfauthor={Changkai Zhang}}

\usepackage{geometry}
\newgeometry{left=0.6in,right=0.6in,top=0.9in,bottom=0.7in}
\usepackage{fancyhdr}

\fancypagestyle{first}{%
    \lhead{}\rhead{}
    \chead{\large PREPRINT FOR PHYSICAL REVIEW (2023)}
}

\usepackage{CJKutf8}
\newcommand{\zh}[1]{\begin{CJK}{UTF8}{gbsn}#1\end{CJK}}
\usepackage{times}
\newlength\mylen
\setcitestyle{citesep={,\kern-\mylen\kern1pt}}
\DeclareUnicodeCharacter{2212}{-}
\newcommand{\prlsec}[1]{\ifdefined\Publication\emph{#1.}\hspace{2pt}---\hspace{2pt}\else\textit{#1.}~---~\fi}

\begin{document}

\preprint{APS/123-QED}


\ifdefined\Publication
\title{\vspace*{12pt}Frustration-Induced Superconductivity in the $\tnn$-$\tnnn$ Hubbard Model \vspace{3pt}}%
\else
\title{\vspace*{12pt}Frustration Induced Superconductivity in the $\tnn$-$\tnnn$ Hubbard Model \vspace{3pt}}%
\fi

\author{Changkai Zhang (\zh{张昌凯})}
\author{Jheng-Wei Li}
\author{Jan von Delft}
\affiliation{Arnold Sommerfeld Center for Theoretical Physics,
Ludwig-Maximilians-Universität München, 80333 Munich, Germany \vspace{5pt}}

\date{\today}

\begin{abstract}
\setstretch{1.08}
The two-dimensional (2D) Hubbard model is widely believed to capture key ingredients of high-$T_c$ superconductivity in cuprate materials. However, compelling evidence remains elusive. In particular, various magnetic orders may emerge as strong competitors of superconducting orders. Here, we study the ground state properties of the doped 2D $\tnn$-$\tnnn$ Hubbard model on a square lattice via the infinite Projected Entangled-Pair State (iPEPS) method with $\mathrm{U}(1)$ or $\mathrm{SU}(2)$ spin symmetry. 
The former is compatible with antiferromagnetic orders, while the latter forbids them. Therefore, we obtain by comparison a detailed understanding of the magnetic impact on superconductivity.
Moreover, an additional $\tnnn$ term accommodates the particle-hole asymmetry, which facilitates studies on the discrepancies between electron- and hole-doped systems.
We demonstrate that (i) a positive $\tnnns/\tnn$ significantly amplifies the strength of superconducting orders; 
(ii) at sufficiently large doping levels, the $\tnn$-$\tnnn$ Hubbard model favors a uniform state with superconducting orders instead of stripe states with charge and spin modulations; and (iii) \ifdefined\Publication the \fi enhancement of magnetic frustration\ifdefined\Publication, by increasing \else~via augmenting \fi either the strength of NNN interactions or the charge doping\ifdefined\Publication, \else ~\fi impairs stripe orders and helps stabilize superconductivity.
\vspace{24pt}
\end{abstract}

\maketitle

\vspace{-2em}

\thispagestyle{first}


\setstretch{0.99}


\noindent\prlsec{Introduction}Despite continuous efforts during the past few decades, the physics of high-$T_c$ superconductivity in cuprate materials \cite{Bednorz&Mueller1986-SC-LaBaCuO} remains unclear. \cite{Keimer-highTc-Review,XGWen2006-highTc-Review} The two-dimensional (2D) Hubbard model \cite{Hubbard1967} on a square lattice is believed to capture the essential low-energy features of cuprates. Various numerical methods \cite{White1992-DMRG,Blankenbecler&Scalapino1981-AFQMC,Sugiyama1986-AFQMC,Knizia&Chan2012-DMET,Cirac2004-PEPS,Cirac2008-PEPS} have been used to tackle this issue. Nevertheless, previous computational attempts generate numerous candidate ground states \cite{NNHubbard-Conclusive,2DHubbard-Benchmark} very close in energies with abundant combinations of charge and spin orders. 
Experiments \cite{Tranquada1995-SC-Period4,Tranquada1996-SC,Tranquada2006-SC-Review,Ghiringhelli2012-SC-CDW,Comin2015-SC-CDW,Wu2011-SC-CDW,Wu2015-SC-CDW,Mesaros2016-SC-CDW} also confirm simultaneous charge and spin modulated states coexisting or competing with superconductivity. This triggers our curiosity on the interplay between the antiferromagnetic (AFM) background and the high-$T_c$ superconductivity in cuprates.

Typical candidates encompass a uniform state \cite{Giamarchi1991-Hubbard-MC,Dagotto1994-Hubbard-Review,Halboth2000-Hubbard-DMRG,Maier2005-Hubbard-MC,Capone2006-Hubbard-DMFT,Eichenberger2007-Hubbard-VWF,Aichhorn2007-Hubbard-cluster-PS,Tocchio2008-Hubbard-PWF,Kancharla2008-Hubbard-cluster,Sordi2012-Hubbard-DMFT,Gull2012-Hubbard-cluster,Yokoyama2012-Hubbard-MC,Kaczmarczyk2013-Hubbard-GWF,Gull2013-Hubbard-Summary,Chen2013-Hubbard-MC,Otsuki2014-Hubbard-DMFT,Deng2015-Hubbard-MC,Tocchio2016-Hubbard-MC} and various stripe states \cite{Poilblanc1989-Hubbard-theory,Zaanen1989-Hubbard-theory,White2003-Hubbard-DMRG,Hager2005-Hubbard-DMRG,Chang2009-Hubbard-AFQMC,Kaczmarczyk2013-Hubbard-GWF,Zhao2017-Hubbard-VMC,NNHubbard-Conclusive,Huang2018-Hubbard-MC&DMRG,Darmawan2018-Hubbard-MC,Vanhala2018-Hubbard-DMFT,Ido2018-Hubbard-MC,Tocchio-Hubbard-JSWF}. The former features a uniform charge density and is commonly associated with $d$-wave superconductivity, while the latter often exhibit charge-density and spin-density waves with diverse periods, with only part of them displaying coexisting superconductivity. For the nearest neighbor (NN) minimal Hubbard model, a series of advanced numerical methods reached a consensus \cite{NNHubbard-Conclusive} that the ground state at $1/8$ hole doping is a filled (one hole per unit cell of the charge order) period 8 stripe state devoid of superconducting orders. The half-filled period 4 stripe state \cite{Tranquada1995-SC-Period4,Mesaros2016-SC-CDW,Tranquada1996-SC-Period4} favored more in, e.g., LaSrCuO materials emerges primarily with negative next-nearest neighbor (NNN) hopping, as demonstrated in numerous computational simulations \cite{Corboz2019-Hubbard-PEPS,Eder1997-ED-tJ,Degotto1999-ED-tJ,Ido2018-Hubbard-MC,HCJiang2019-Hubbard-DMRG,Zheng2016-DMET-Hubbard,White&Schollwoeck2020-DMRG-Hubbard-PlaquettePairing,Jiang&Devereaux2020-Hubbard-DMRG}.
This motivates our investigations beyond the minimal Hubbard model.

Concurrently, multiple recent studies \cite{SSGong2021-tJ-DMRG,STJiang&Scalapino&White2021-t1t2J,STJiang&Scalapino&White2022-tttJ} focusing on the extended $t$-$J$ model have uncovered substantially more robust superconducting orders in electron-doped settings as opposed to hole-doped configurations, a finding that contradicts experimental observations.
\ifdefined\Publication \hspace{-3pt}Explorations of the extended Hubbard model using Density Matrix Renormalization Group (DMRG) have yielded inconsistent outcomes \cite{Jiang&Devereaux2023-Hubbard-ehdoped,Schollwoeck&White2023-Hubbard-ehdoped}, further underscoring the significance of researches beyond the minimal Hubbard model.
\else Explorations of the extended Hubbard model through a range of numerical techniques yields inconsistent outcomes \cite{Jiang&Devereaux2023-Hubbard-ehdoped,Schollwoeck&White2023-Hubbard-ehdoped}, further underscoring the significance of pursuing research beyond the minimal Hubbard model. \fi


In this paper, we use the infinite Projected Entangled-Pair State (iPEPS) \cite{Cirac2004-PEPS,Cirac2008-PEPS} ansatz and simple update algorithm \cite{HCJiang2008-SimpleUpdate} to study the ground state properties of the $\tnn$-$\tnnn$ Hubbard model. \ifdefined\Publication Our iPEPS is less susceptible to finite-size effects than DMRG on cylinders. Leveraging \else Specifically, leveraging \fi our cutting-edge QSpace tensor library \cite{Weichselbaum2012-QSpace,Weichselbaum2012-QSpace-XSymbols}, we are capable of conducting simulations with $\mathrm{U}(1)$ or $\mathrm{SU}(2)$ spin symmetry, where the former admits local magnetic moments and the latter forbids them. This allows us to scrutinize the impact of magnetic orders on pairing properties.~Our simulations demonstrate that (i) a positive $\tnnns/\tnn$ significantly amplifies the strength of superconducting orders; 
\ifdefined\Publication (ii) at sufficiently large doping, the $\tnn$-$\tnnn$ Hubbard model favors an $\mathrm{SU}(2)$ uniform state with $d$-wave pairing orders instead of a $\mathrm{U}(1)$ stripe state in \cite{Corboz2019-Hubbard-PEPS}; \else (ii) at sufficiently large doping levels, the $\tnn$-$\tnnn$ Hubbard model favors an $\mathrm{SU}(2)$ symmetric uniform state with $d$-wave pairing orders instead of a $\mathrm{U}(1)$ symmetric stripe state observed in previous research \cite{Corboz2019-Hubbard-PEPS}; \fi and (iii) \ifdefined\Publication the \fi enhancement of magnetic frustration\ifdefined\Publication, by increasing \else via augmenting \fi either the strength of NNN interactions or the charge doping\ifdefined\Publication, \else ~\fi impairs stripe orders and helps stabilize superconductivity.


\vspace*{7pt}

\noindent\prlsec{Model}The 2D $\tnn$-$\tnnn$ Hubbard model on a square lattice is defined via the following Hamiltonian
\begin{equation}\label{Hamiltonian}
    \mathcal{H} = -\sum_{i,j,\sigma} t_{ij} \left[\, c^\dagger_{i\sigma} c_{j\sigma} + \text{h.c.} \,\right] + U\sum_i n_{i\uparrow} n_{i\downarrow}.
    \vspace{-4pt}
\end{equation}
\noindent Here, $t_{ij} = \tnn$ or $\tnnn$ for NN or NNN, respectively, and zero otherwise; $U$ measures the on-site Coulomb repulsion. Throughout this paper, we use $U/\tnn = 10$, as established to be realistic for cuprate materials \cite{Hirayama2018-Hubbard-parameter,Hirayama2018-Hubbard-electronic}, and set $\tnn=1$ for convenience.

\begin{figure*}[htp!]
    \ifdefined\Publication\vspace{7pt}\else\vspace{11pt}\fi
    \centering
    \hspace*{-20pt}
    \ifdefined\Publication
    \includegraphics[width=0.95\textwidth]{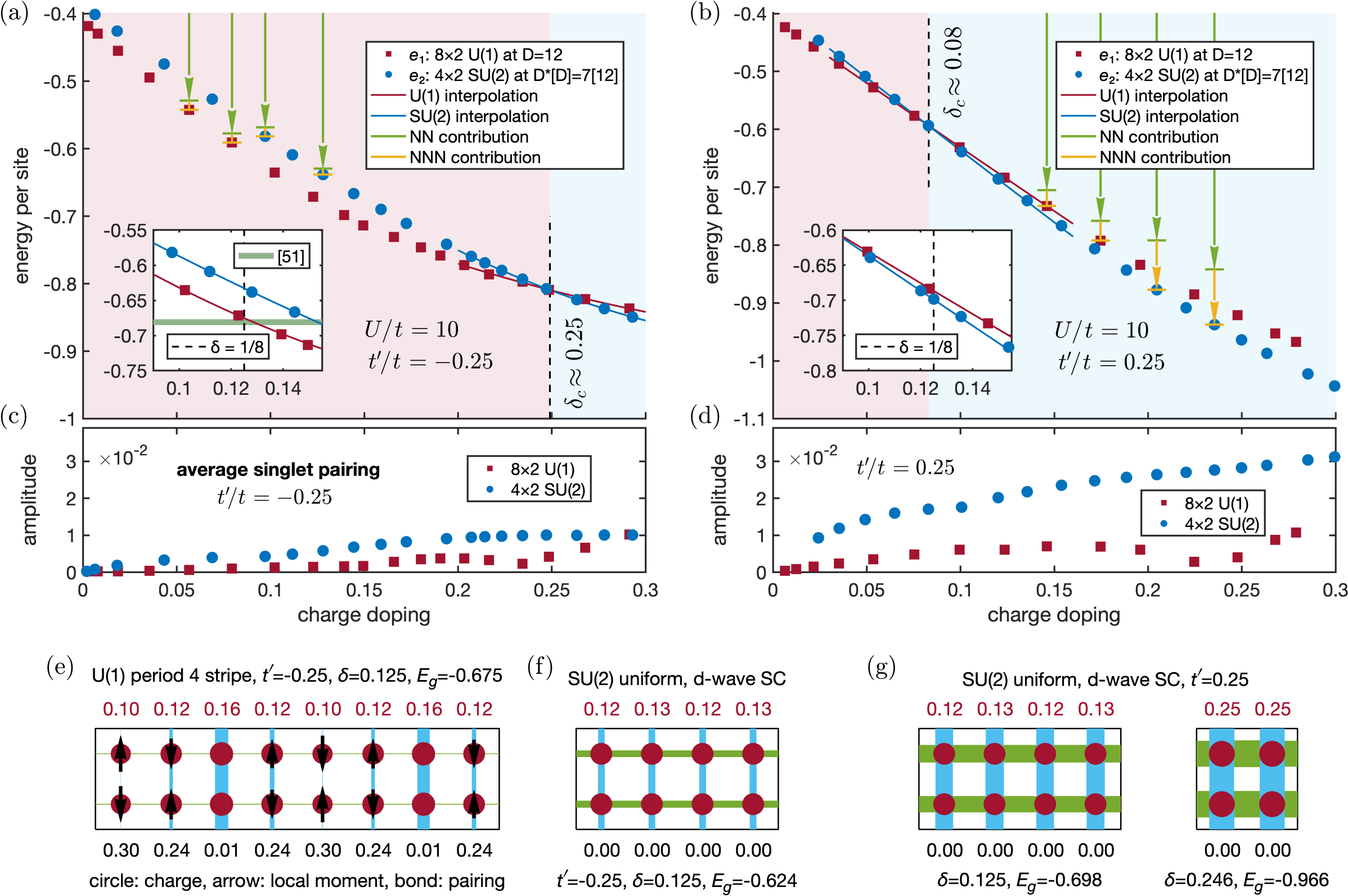}
    \else\includegraphics[width=0.95\textwidth]{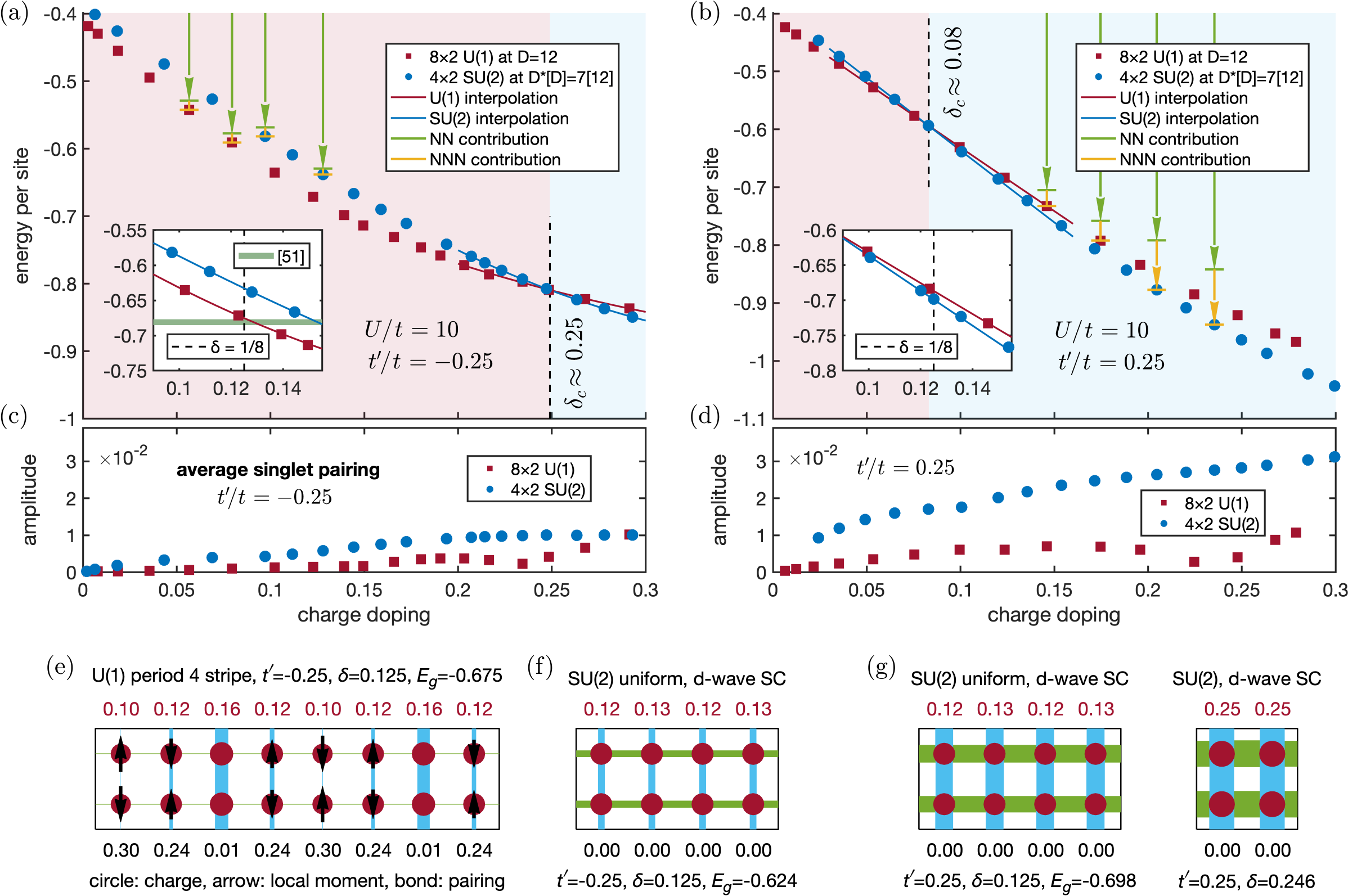}\\ \fi
    \begin{minipage}{0.9\textwidth}
    \vspace{4pt}
    \caption{The ground state energy per site (a,b) and singlet pairing (c,d) vs.~doping $\delta$ of the $\tnn$-$\tnnn$ Hubbard model at $U/\tnn\!=\!10$ and (a,c) $\tnnns/\tnn\!=\!-0.25$ or (b,d) $\tnnns/\tnn\!=\!0.25$, computed via $\mathrm{U}(1)$ iPEPS (red squares) on an 8×2 supercell at bond dimension $D\!=\!12$ and $\mathrm{SU}(2)$ iPEPS (blue circles) on a 4×2 supercell keeping $D^*\!=\!7$ multiplets (bond dimension $D\!=\!12$). Green and yellow arrows, respectively, indicate the NN (including on-site) and NNN contributions to the energy for several typical data points. Inset: zoom into the region near $1/8$ doping. (e-g): Details of the $\mathrm{U}(1)$ and $\mathrm{SU}(2)$ symmetric ground states on 8×2, 4×2 and 2×2 supercells. Areas of red circles and lengths of black arrows are proportional to the charge density (top rows) and the local moments (bottom rows), respectively. Bond widths indicate NN singlet pairing amplitudes and two different colors indicate opposite signs. \ifdefined\Publication For (f-g), we used $D^*[D]\!=\!8[13]$ for reasons explained in the Supplemental Material.\vspace{-5pt}\fi}
    \label{Energy}
    \end{minipage}
\end{figure*}


\vspace*{7pt}

\noindent\prlsec{Method}In our computations, we apply the fermionic iPEPS \cite{Corboz2010-PEPS-NN,Corboz2010-PEPS-NNN,Bruognolo2020-iPEPS-Review,Cirac2010-fermionicPEPS,Barthel2009-PEPS-Contract,Cirac2008-Spin-PEPS&DMRG} ansatz, a tensor network method targeting 2D lattice models, to simulate the $\tnn$-$\tnnn$ Hubbard model in the thermodynamic limit. The ansatz exploits translational symmetry by assuming that the infinite tensor network consists of periodically repeated supercells of tensors. Each supercell comprises several rank-5 tensors with one physical index carrying states in the local Hilbert space
, and four auxiliary indices connecting neighboring sites. The accuracy of the simulation can be controlled \ifdefined\Publication \else systematically \fi by the bond dimensions of the auxiliary indices. Different supercell sizes yield stripe states with different periods in charge or spin orders. Previous researches \cite{Corboz2019-Hubbard-PEPS,Jiang&Devereaux2023-Hubbard-ehdoped,SSGong2021-tJ-DMRG} on the Hubbard model or the $t$-$J$ model have identified stripe states with period 4 charge orders as a representative stripe state. Therefore, we hereby focus on the period 4 stripe state. Further discussions \ifdefined\Publication and details \fi regarding stripes with longer periods can be found in the Supplemental Material.

The optimization is performed via imaginary time evolution \cite{Vidal2007-iTEBD} in which projector $\exp\{-\tau(\mathcal{H}+\mu N)\}$ ($\tau$ is a small number, $\mathcal{H}$ the Hamiltonian, $\mu$ the chemical potential and $N$ the charge density) is repeatedly applied to some random initial state until the ground state energy converges. Models with NNN interactions are computationally very expensive. Therefore, we choose the simple update scheme \cite{HCJiang2008-SimpleUpdate,Corboz2010-PEPS-NN,Corboz2010-PEPS-NNN} for a balance between accuracy and computational complexity.~Observables are extracted by contracting the tensor network using corner transfer matrix method \cite{Corboz2010-PEPS-NN,Corboz&Vidal2011-tJ-PEPS&DMRG,Nishino1996-CTMRG,Corboz2014-tJ-PEPS,Nishino1996-CTMRG,Vidal2009-CTMRG,Bruognolo2020-iPEPS-Review}. The QSpace tensor library \cite{Weichselbaum2012-QSpace,Weichselbaum2012-QSpace-XSymbols,osqspacev4} is used to implement either $\mathrm{U}(1)$ or $\mathrm{SU}(2)$ spin symmetry.

The $\mathrm{U}(1)$ iPEPS simulations are conducted on an 8×2 supercell at bond dimension $D\!=\!12$. This is required for capturing the period 4 charge orders, as the corresponding spin order periods are typically twice as long as charge periods. The $\mathrm{SU}(2)$ iPEPS simulations are performed on a 4×2 or 2×2 supercell by keeping $D^*\!=\!7$ symmetry multiplets (corresponding to a bond dimension $D\!=\!12$) \cite{Weichselbaum2012-QSpace}. Spin orders are suppressed upon enforcing $\mathrm{SU}(2)$ symmetry, making a 4×2 supercell adequate to detect any potential period 4 orders, while the 2×2 supercell is employed to ascertain the uniformity of the ground state. Charge doping is adjusted by tuning the chemical potential.

\begin{figure}
    \vspace{10pt}
    \centering
    \ifdefined\Publication
    \includegraphics[width=0.47\textwidth]{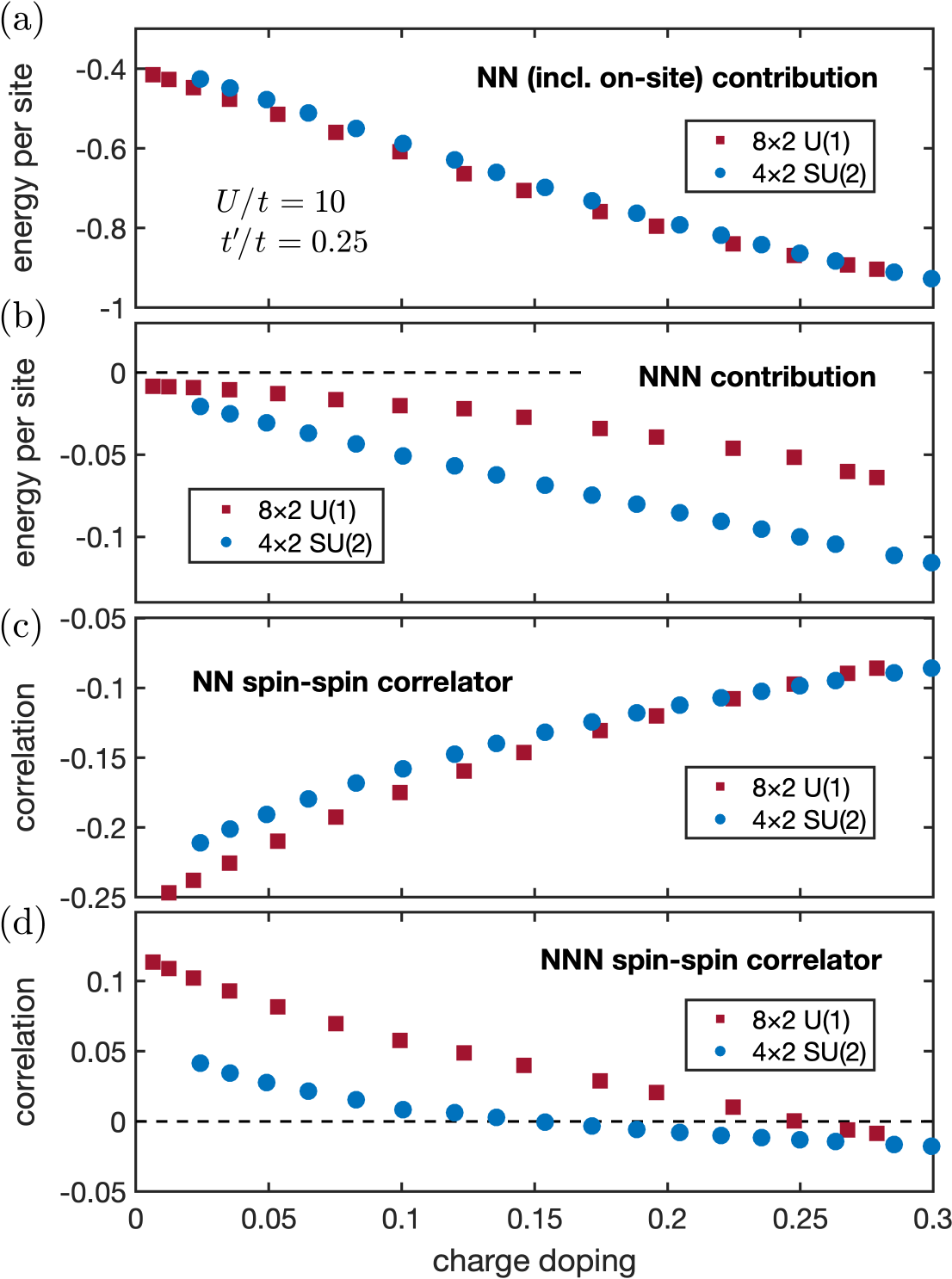}
    \else\includegraphics[width=0.47\textwidth]{img/Measure.pdf}\\ \fi
    \begin{minipage}{0.45\textwidth}\vspace{3pt}
    \caption{The contribution of (a) the NN (including on-site) and (b) the NNN terms to the total energy per site in the $\mathrm{U}(1)$ and $\mathrm{SU}(2)$ ground states, respectively, as a function of doping. (c) The NN and (d) the NNN spin-spin correlators in the $\mathrm{U}(1)$ and $\mathrm{SU}(2)$ ground states, respectively.}
    \label{Energetics}
    \end{minipage}
    \vspace{-7pt}
\end{figure}


\vspace{7pt}

\noindent\prlsec{Energetics}Figures~\ref{Energy}(a,b) shows the ground state energy per site of the $\tnn$-$\tnnn$ Hubbard model as a function of doping under $U/\tnn\!=\!10$ and $\tnnns/\tnn\!=\!\mp0.25$, computed via the $\mathrm{U}(1)$ and $\mathrm{SU}(2)$ iPEPS and denoted as $\eA$ (red) and $\eS$ (blue), respectively. Figures~\ref{Energy}(c,d) shows the corresponding singlet pairing amplitudes. Figures~\ref{Energy}(e) and \ref{Energy}(f) display, respectively, the detailed characteristics of the $\mathrm{U}(1)$ and $\mathrm{SU}(2)$ ground states with a negative $\tnnns/\tnn$ at the predominantly studied $1/8$ doping. Figure~\ref{Energy}(g) presents $\mathrm{SU}(2)$ ground states with a positive $\tnnns/\tnn$ showcasing numerically significant $d$-wave singlet pairing orders. 

Utilizing an 8×2 supercell, our $\mathrm{U}(1)$ iPEPS generates a non-superconducting stripe state with a period 4 charge density wave and a period 8 antiferromagnetically ordered spin density wave. These attributes, along with the ground state energy acquired, are generally consistent with the findings in \cite{Corboz2019-Hubbard-PEPS}. By contrast, when we enforce the $\mathrm{SU}(2)$ symmetry and suppress spin orders, we find a uniform state without any charge orders, at odds with finite-size studies \cite{HCJiang2019-Hubbard-DMRG,Jiang&Devereaux2023-Hubbard-ehdoped,SSGong2021-tJ-DMRG}. Moreover, strong $d$-wave pairing emerges for positive $\tnnns/\tnn$\ifdefined\Publication, \else~\fi which implies superconductivity. $\mathrm{SU}(2)$ iPEPS on 4×2 and 2×2 supercells produce physically identical states, confirming the uniformity of the ground state. 

Near zero doping, we find $\eS\!>\!\eA$. This is consistent with the well-established fact that the Heisenberg model on a square lattice has an AFM ground state which breaks $\mathrm{SU}(2)$ symmetry. However, as the doping increases, $\eS$ decreases faster than $\eA$. They intersect at $\delta_c\approx0.25$ for $\tnnns/\tnn\!=\!-0.25$ and $\delta_c\approx0.08$ for $\tnnns/\tnn\!=\!0.25$, as depicted in Fig.~\ref{Energy}(a,b), in agreement with prior observations \cite{Corboz2019-Hubbard-PEPS} that a \ifdefined\Publication negative or positive $\tnnns/\tnn$ favors stripe or uniform states, respectively. \else negative $\tnnns/\tnn$ encourages striped states whereas a positive $\tnnns/\tnn$ facilitates uniform states. \fi Intuitively, a positive $\tnnns/\tnn$ promotes diagonal hopping of the doped charges, which in turn disrupts the AFM background in the vicinity of the domain wall within the stripe states, rendering the presence of domain walls less desirable\ifdefined\Publication~\cite{Huang2018-Hubbard-MC&DMRG}. \else. \cite{Huang2018-Hubbard-MC&DMRG}\fi

The lower energy of the $\mathrm{SU}(2)$ \ifdefined\Publication relative to the $\mathrm{U}(1)$ \fi ground state at large doping can be understood as the result of magnetic frustration induced by the NNN hoppings. The $\mathrm{U}(1)$ stripe state still accommodates AFM orders and thus suffers strongly from magnetic frustrations with NNN hopping. \ifdefined\Publication By contrast, \else On the contrary, \fi the $\mathrm{SU}(2)$ uniform state is less frustrated since it hosts no local spin orders. Indeed, the NNN terms contribute much less to lowering the energy $\eA$ of the stripe state than to the energy $\eS$ of the uniform state, as indicated via the yellow arrows in Fig.~\ref{Energy}(a,b).  

This issue is further elaborated in Figs.~\ref{Energetics}(a) and \ref{Energetics}(b), showing the contribution of NN (including on-site) and NNN terms to the total energy per site as a function of doping, respectively.
Throughout the entire doping range in our study, the NN contribution is marginally lower in the $\mathrm{U}(1)$ states than in the $\mathrm{SU}(2)$ states. Conversely, the NNN contribution is substantially greater in the $\mathrm{U}(1)$ \ifdefined\Publication than \else states compared to \fi the $\mathrm{SU}(2)$ cases, ultimately leading to a lower overall energy for the $\mathrm{SU}(2)$ states at large doping levels.~As a comparison, Figs.~\ref{Energetics}(c) and \ref{Energetics}(d) show the NN and NNN spin-spin correlators, respectively.~The NN correlations stay negative for both $\mathrm{U}(1)$ and $\mathrm{SU}(2)$ states, reflecting the overall AFM background. The NNN correlations, however, turn negative considerably sooner for the $\mathrm{SU}(2)$ states than for the $\mathrm{U}(1)$ states, echoing the findings in ultracold atom experiments that doped charges drive the NNN spin-spin correlation negative
\cite{Koepsell2021-Ultracold-Polaron,Koepsell2019-Ultracold-FermiHubbard,Koepsell2020-Ultracold-tech,Grusdt2019-Hubbard-StringPattern,Chen&vonDelft2021-Hubbard-XTRG}. This indicates that the $\mathrm{SU}(2)$ state better reconciles the magnetic frustration, thereby achieving a lower NNN energy. Such behaviors exemplify how the enhancement of magnetic frustration through NNN hopping inhibits the formation of stripes and promotes the emergence of superconductivity.


\begin{figure*}[htp!]
    \vspace{-2pt}
    \centering
    \hspace{-5pt}
    \ifdefined\Publication
    \subfloat{\includegraphics[width=0.46\textwidth]{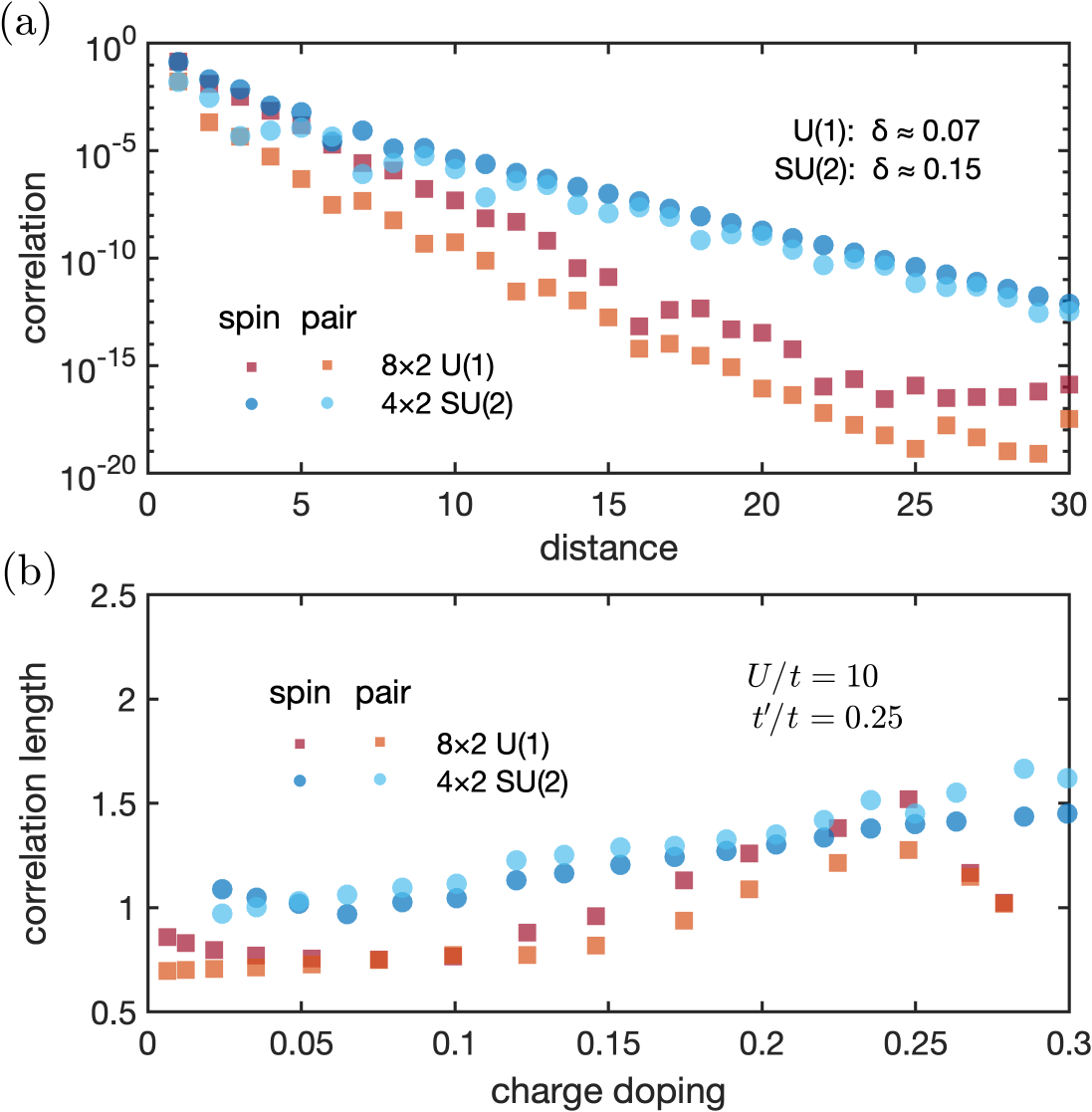}}
    \else\subfloat{\includegraphics[width=0.46\textwidth]{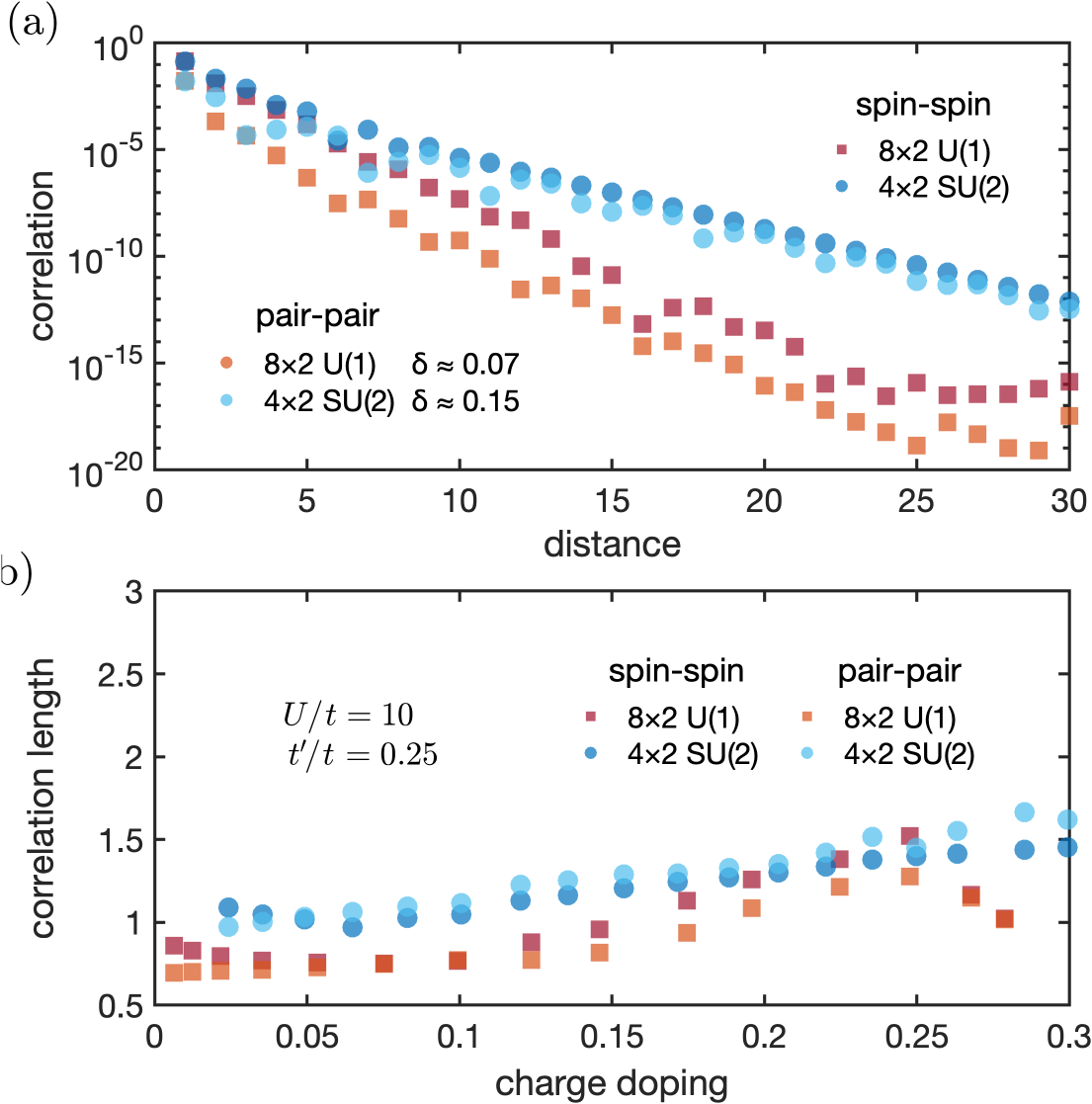}}\fi
    \hspace{0.062\textwidth}
    \ifdefined\Publication
    \subfloat{\includegraphics[width=0.446\textwidth]{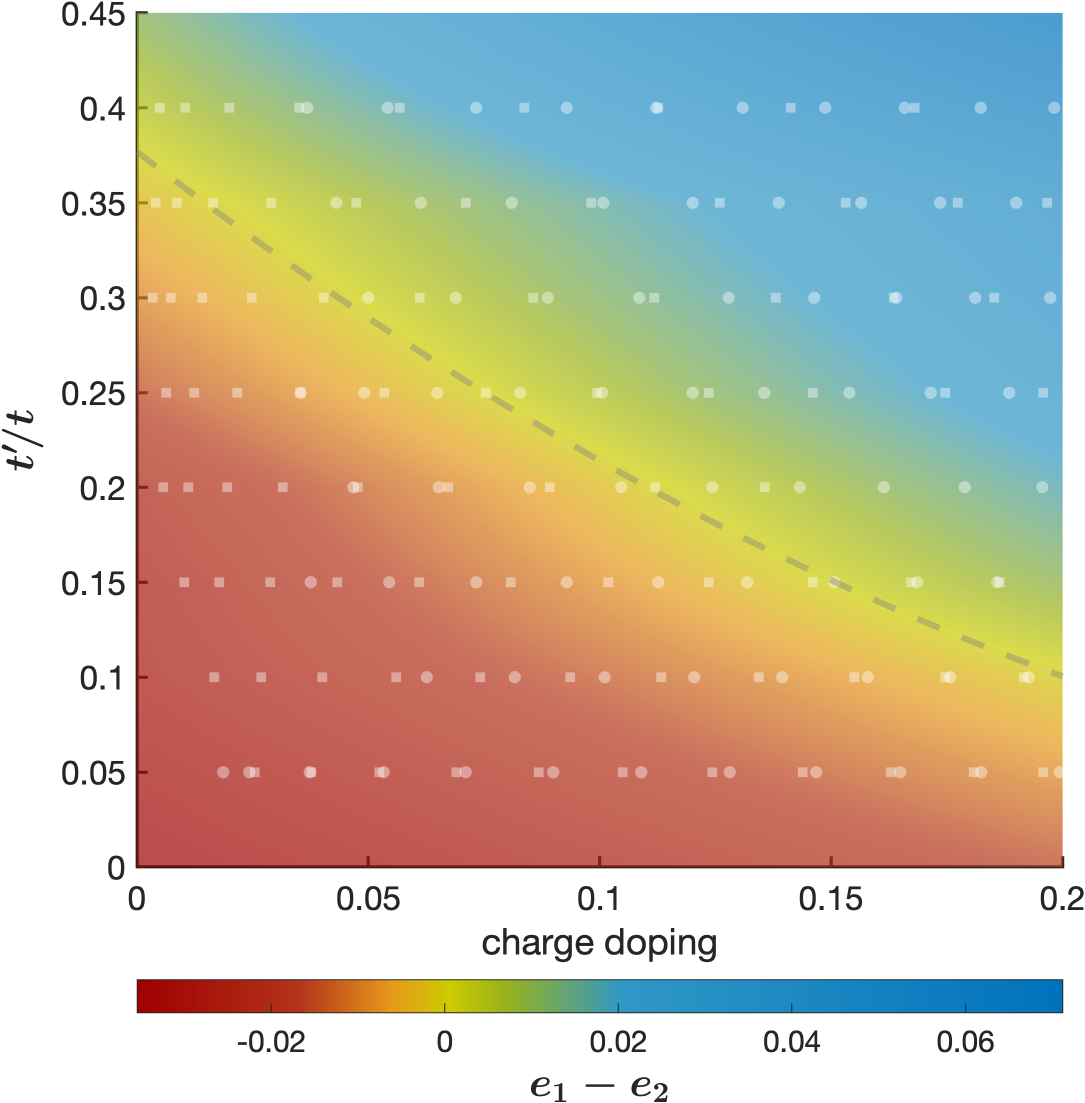}}\\
    \else\subfloat{\includegraphics[width=0.446\textwidth]{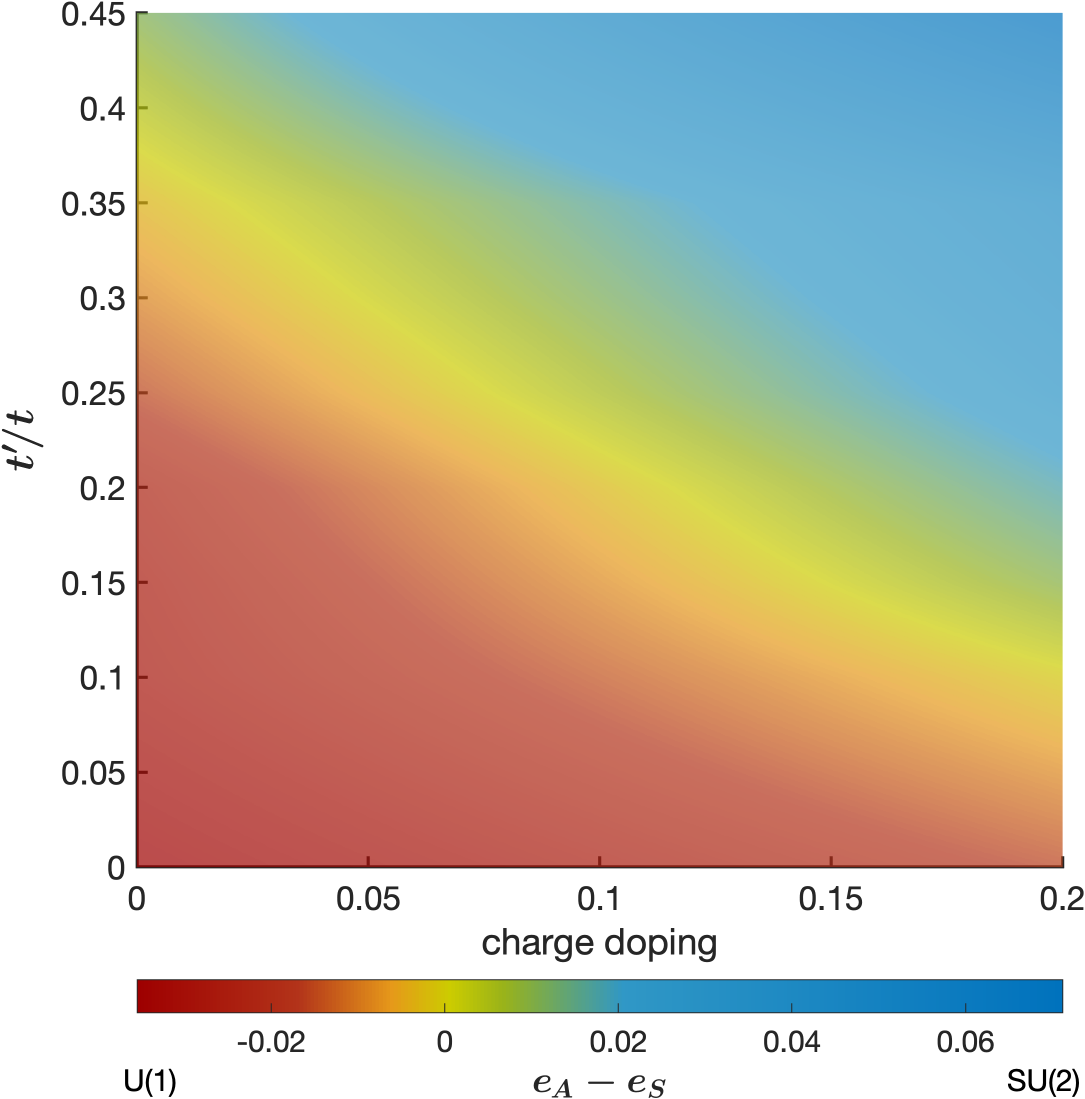}}\\ \fi
    \hspace{1pt}
    \begin{minipage}{0.44\textwidth}
    \caption{(a) The long-range spin-spin and pair-pair correlators in the $\mathrm{U}(1)$ and $\mathrm{SU}(2)$ ground states, respectively. All these correlators exhibit an exponential decay behavior. (b) The corresponding correlation lengths as a function of doping.}
    \label{Correlation}
    \end{minipage}
    \hspace{0.062\textwidth}
    \begin{minipage}{0.44\textwidth}
    \caption{\ifdefined\Publication The ground state phase diagram of the $\tnn$-$\tnnn$ Hubbard model with respect to doping and $\tnnns/\tnn$. The color scale indicates $\eA - \eS$, obtained via linear interpolation from a discrete set of scanning points (white). The grey dashed line marks $\eA = \eS$. \else The ground state \emph{phase diagram} of the $\tnn$-$\tnnn$ Hubbard model with respect to doping and NNN hopping amplitude. The color scale indicates $\eA - \eS$. Red/blue color marks $\mathrm{U}(1)$/$\mathrm{SU}(2)$ region, with the \emph{phase boundary} highlighted by the yellow strip.\fi}
    \label{Phase}
    \end{minipage}
    \vspace{-1pt}
\end{figure*}

\vspace{8pt}

\noindent\prlsec{Pairing Order}The superconducting order can be characterized by the singlet pairing amplitude $\Delta_{\mathbf{r},\mathbf{s}} = \langle c_{\mathbf{r}\uparrow} c_{\mathbf{s}\downarrow} - c_{\mathbf{r}\downarrow} c_{\mathbf{s}\uparrow} \rangle$. Specifically, we focus on the NN singlet pairing. As illustrated in Figs.~\ref{Energy}(e-g), we observe finite singlet pairing orders for both $\mathrm{U}(1)$ and $\mathrm{SU}(2)$ ground states. However, the pairing amplitude (averaged over the supercell) of the $\mathrm{SU}(2)$ states can be substantially larger than that in the $\mathrm{U}(1)$ states throughout the entire doping range for positive $\tnnns/\tnn$, as presented in Fig.~\ref{Energy}(d). This can be attributed to the fact that the $\mathrm{SU}(2)$ iPEPS is, by construction, a spin-singlet state\ifdefined\Publication. Indeed, the latter \else~which \fi can be interpreted as a generalized version of the resonating valence bond (RVB) state\ifdefined\Publication~\cite{JWLi2021-tJ-PEPS}. \else. \cite{JWLi2021-tJ-PEPS} \fi Therefore, the existence of $d$-wave pairing order is reminiscent of Anderson's original RVB proposal \cite{Anderson1987-RVB,Anderson1987-RVBSC}.

Moreover, we discover that the singlet pairing for positive $\tnnns/\tnn$ can be considerably larger than that for negative $\tnnns/\tnn$. Intuitively, this could be perceived as pair formation being enhanced (reduced) by the constructive (destructive) interference between NN and NNN hopping at positive (negative) $\tnnns/\tnn$\ifdefined\Publication~\cite{Dagotto2001-QualitativeNNN}. \else.~\cite{Dagotto2001-QualitativeNNN} \fi This is in line with prior findings in the extended $t$-$J$ model \cite{SSGong2021-tJ-DMRG,STJiang&Scalapino&White2021-t1t2J,STJiang&Scalapino&White2022-tttJ} and Hubbard model \cite{Jiang&Devereaux2023-Hubbard-ehdoped} using Density Matrix Renormalization Group. Electronic structure analysis \cite{Hirayama2018-Hubbard-parameter,Hirayama2018-Hubbard-electronic,Tohyama&Maekawa1994-t2sign,Andersen&Liechtenstein1995-LDA-DFT} \ifdefined\Publication suggests \else reveals \fi that positive (negative) $\tnnns/\tnn$ corresponds to electron- (hole-) doped cuprates. Consequently, the numerics so far yield outcomes that are opposite to the experimental observations, where hole-doped cuprates exhibit stronger superconductivity. This emphasizes the necessity for further investigations regarding the appropriate parameter settings in the effective models \cite{STJiang&Scalapino&White2021-t1t2J,Xiang2009-ElectronDopedCuprates,Jiang&Scalapino&White2023-Hubbard-3to1}.

\vspace{8pt}

\noindent\prlsec{Long-range Order}Figure~\ref{Correlation}(a) displays the long-range spin-spin $S_{\mathbf{r}\mathbf{s}}\! = \!\langle \mathbf{S}_\mathbf{r}\cdot \mathbf{S}_\mathbf{s}\rangle - \langle \mathbf{S}_\mathbf{r}\rangle\cdot\langle\mathbf{S}_\mathbf{s}\rangle$ and pair-pair $P_{\mathbf{r}\mathbf{s}} = \langle\Delta^\ey_{\mathbf{r}}\Delta^\ey_{\mathbf{s}}\rangle - \langle\Delta^\ey_{\mathbf{r}}\rangle\langle\Delta^\ey_{\mathbf{s}}\rangle$ (where $\Delta^\alpha_\mathbf{r} = \Delta_{\mathbf{r},\mathbf{r+\alpha}}$ and $\mathbf{\alpha} = \mathbf{x},\mathbf{y}$ is the horizontal or vertical unit vector) correlators for two specific ground states with $\mathrm{U}(1)$ or $\mathrm{SU}(2)$ symmetry for $\tnnns/\tnn>0$. Figure~\ref{Correlation}(b) shows the corresponding correlation lengths. Our data indicate that all these correlators decay exponentially, and the correlation lengths never exceed two units throughout the entire doping range. This suggests no connected long-range spin or pairing orders in both scenarios. Accordingly, a minor local pairing order sufficiently signals weak superconductivity in the stripe states.

\vspace{8pt}

\noindent\prlsec{Phase Diagram}Figure~\ref{Phase} presents a schematic ground state \emph{phase diagram} for $\tnnns/\tnn\!>\!0$ derived via linear interpolation from a discrete set of scanning points. The $\mathrm{U}(1)$ stripe states are energetically favored in the bottom-left corner, and the $\mathrm{SU}(2)$ uniform states the top-right corner. This is generally consistent with previous studies on $\tnn$-$\tnnn$-$J$ model \cite{SSGong2021-tJ-DMRG}. Therefore, an increase of either charge doping or the NNN hopping, which both intensify magnetic frustration, will drive the ground state from striped to uniform states. Recall that the uniform ground states are typically accompanied by strong superconductivity. The phase diagram thus supports the conclusion that the enhancement of magnetic frustration helps stabilize superconductivity.


\vspace{8pt}

\noindent\prlsec{Discussion}In this research, we have studied the ground state properties and the phase diagram of the $\tnn$-$\tnnn$ Hubbard model via $\mathrm{U}(1)$ and $\mathrm{SU}(2)$ symmetric iPEPS method. We discovered an $\mathrm{SU}(2)$ uniform state with strong $d$-wave superconducting orders, with a lower energy than the striped $\mathrm{U}(1)$ states at large doping levels. 
Although the variational space of $\mathrm{U}(1)$ iPEPS is larger than that of $\mathrm{SU}(2)$ iPEPS, the fact that $\mathrm{U}(1)$ iPEPS has so far failed to yield a uniform ground state suggests that $\mathrm{U}(1)$ iPEPS has difficulty handling the subspace devoid of magnetic orders\ifdefined\Publication. \else (without \emph{a priori} guidance about the $\mathrm{SU}(2)$ compatible settings, see the Supplemental Material for more details). \fi This highlights the importance of exploring quantum states with several different global symmetries in tensor network simulations. \ifdefined\Publication We note, however, that it is possible to recover the $\mathrm{SU}(2)$ ground states via a $\mathrm{U}(1)$ implementation with \emph{a priori} guidance about the $\mathrm{SU}(2)$ compatible settings, see the Supplemental Material for more details. \fi

Also, we have demonstrated the interplay between local magnetic orders and superconductivity. The additional NNN interaction terms introduce extra magnetic frustration and help suppress the AFM orders, favoring strong $d$-wave superconductivity at large doping levels. Besides, a positive $\tnnns/\tnn$ frustrates the domain walls and stimulates pair formation. This suggests that the superconductivity in cuprate materials can be enhanced, and $T_c$ incremented, by elevating the strength of NNN hopping.

\vspace{8pt}

\noindent\prlsec{Outlook}The novel $\mathrm{SU}(2)$ ground state, expressed in terms of iPEPS tensor network, contains information on dominant contributions from the many-body Hilbert space. Consequently, it is possible to generate various \emph{snapshots} of the type accessible via quantum gas microscopy in the ultracold atom experiments \cite{Koepsell2020-Ultracold-tech,Koepsell2019-Ultracold-FermiHubbard}, enabling a direct comparison with experimental analysis \cite{Chen&vonDelft2021-Hubbard-XTRG,Chen&Li2022-tJ-XTRG}.~Such information would facilitate further investigations regarding the dopant mobility through high-order correlators \cite{Grusdt2021-Ultracold-Correlator,Grusdt&Cirac2020-iPEPS-Correlator} or string patterns using suitable pattern recognition algorithms \cite{Grusdt2019-Hubbard-StringPattern}.~Also, similar $\mathrm{SU}(2)$ symmetric tensor techniques can be applied to some thermal tensor network methods, such as finite temperature PEPS \cite{Czarnik2012-PEPS-finiteT-first,Czarnik2014-PEPS-finiteT-fermionic,Czarnik2015-PEPS-finiteT-variational}, Exponential Tensor Renormalization Group (XTRG) \cite{Li&Weichselbaum2018-XTRG,Li&vonDelft2019-Heisenberg-XTRG,Chen&Li2022-tJ-XTRG} or tangent space Tensor Renormalization Group (tanTRG) \cite{Li2022-tanTRG} to explore physics at finite temperatures where strange metal behavior is observed experimentally.

\vspace{8pt}

\noindent\prlsec{Acknowledgement}We thank our colleagues Andreas Weichselbaum and Andreas Gleis for stimulating discussions and technical suggestions, which lead to a significant speedup of the algorithms. We also thank Philippe Corboz for helpful feedback on a preliminary version of this work, as well as Zi Yang Meng for constructive comments. This research was funded in part by the Deutsche Forschungsgemeinschaft under Germany's Excellence Strategy EXC-2111 (Project No.~390814868), and is part of the Munich Quantum Valley, supported by the Bavarian state government through the Hightech Agenda Bayern Plus.


\bibliography{czhang}

\end{document}


\preprint{APS/123-QED}


\title{Supplemental Material $-$ Frustration Induced Superconductivity in the $\tnn$-$\tnnn$ Hubbard Model \vspace{3pt}}%

\author{Changkai Zhang (\zh{张昌凯})}
\author{Jheng-Wei Li}
\author{Jan von Delft}
\affiliation{Arnold Sommerfeld Center for Theoretical Physics,
Ludwig-Maximilians-Universität München, 80333 Munich, Germany \vspace{5pt}}

\date{\today}

\maketitle

\thispagestyle{first}


\setstretch{1.03}

In the Supplemental Material, we discuss (\ref{RefinedSU}) some technical details for accelerating the optimization of infinite Projected Entangled-Pair State (iPEPS) tensor networks for next-nearest neighbor (NNN) lattice models, (\ref{Technical},\ref{BenchmarkSection}) technical details and benchmarks of the $\mathrm{SU}(2)$ iPEPS, (\ref{Guidance}) guided and unguided $\mathrm{U}(1)$ iPEPS optimization, (\ref{Singlet}) artifacts of singlet pairing, and (\ref{LongStripes}) stripes with longer periods.

\vspace{-3pt}
\section{Refined Simple Update Scheme for Next-nearest Neighbor Models}
\label{RefinedSU}
\vspace{-2pt}

\begin{figure}[htp!]
    \vspace*{0pt}
    \centering
    \hspace*{6em}\includegraphics[scale=0.16]{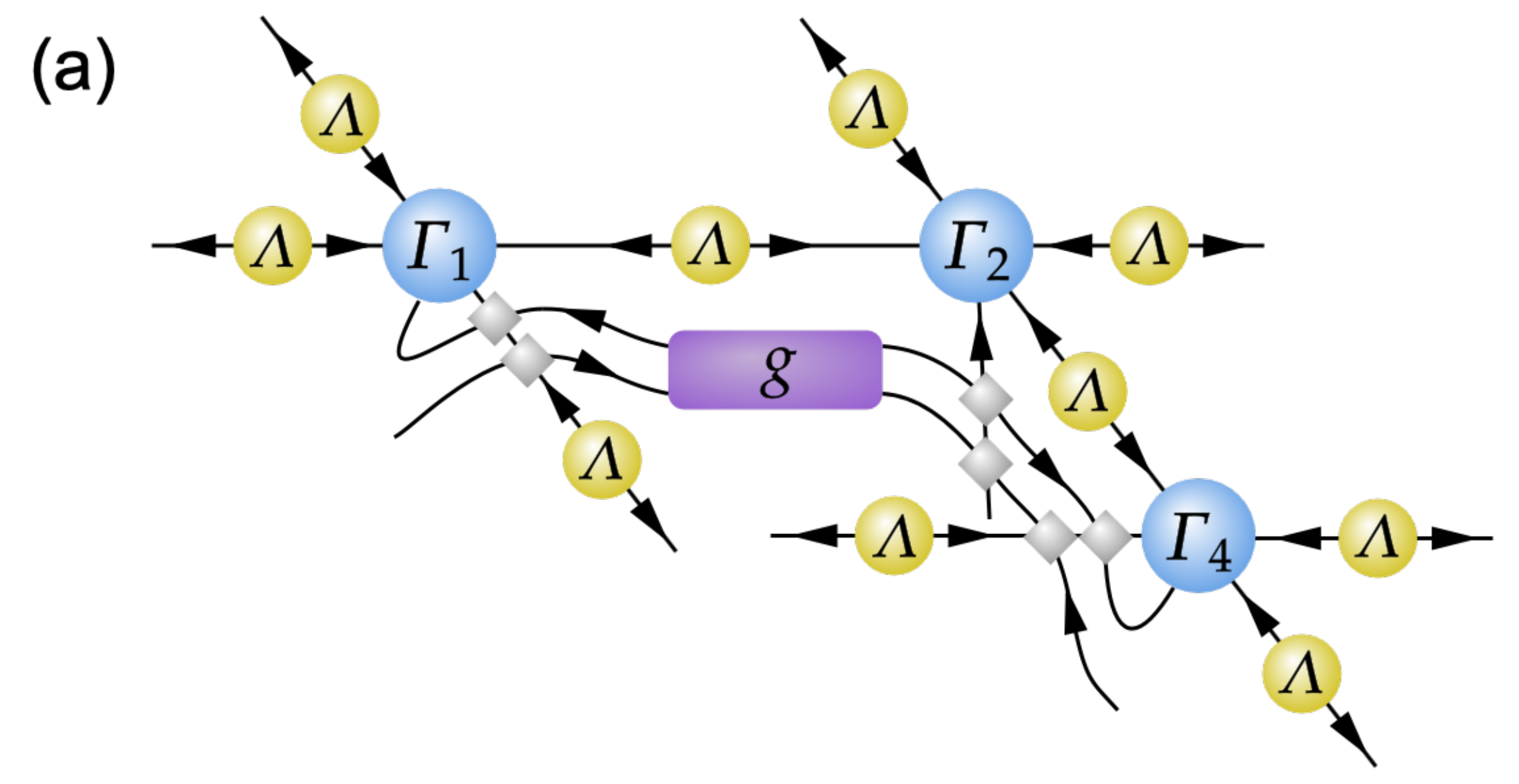}\\[-6.1em]
    \ifdefined\Publication
    \hspace*{-6em}\includegraphics[scale=0.16]{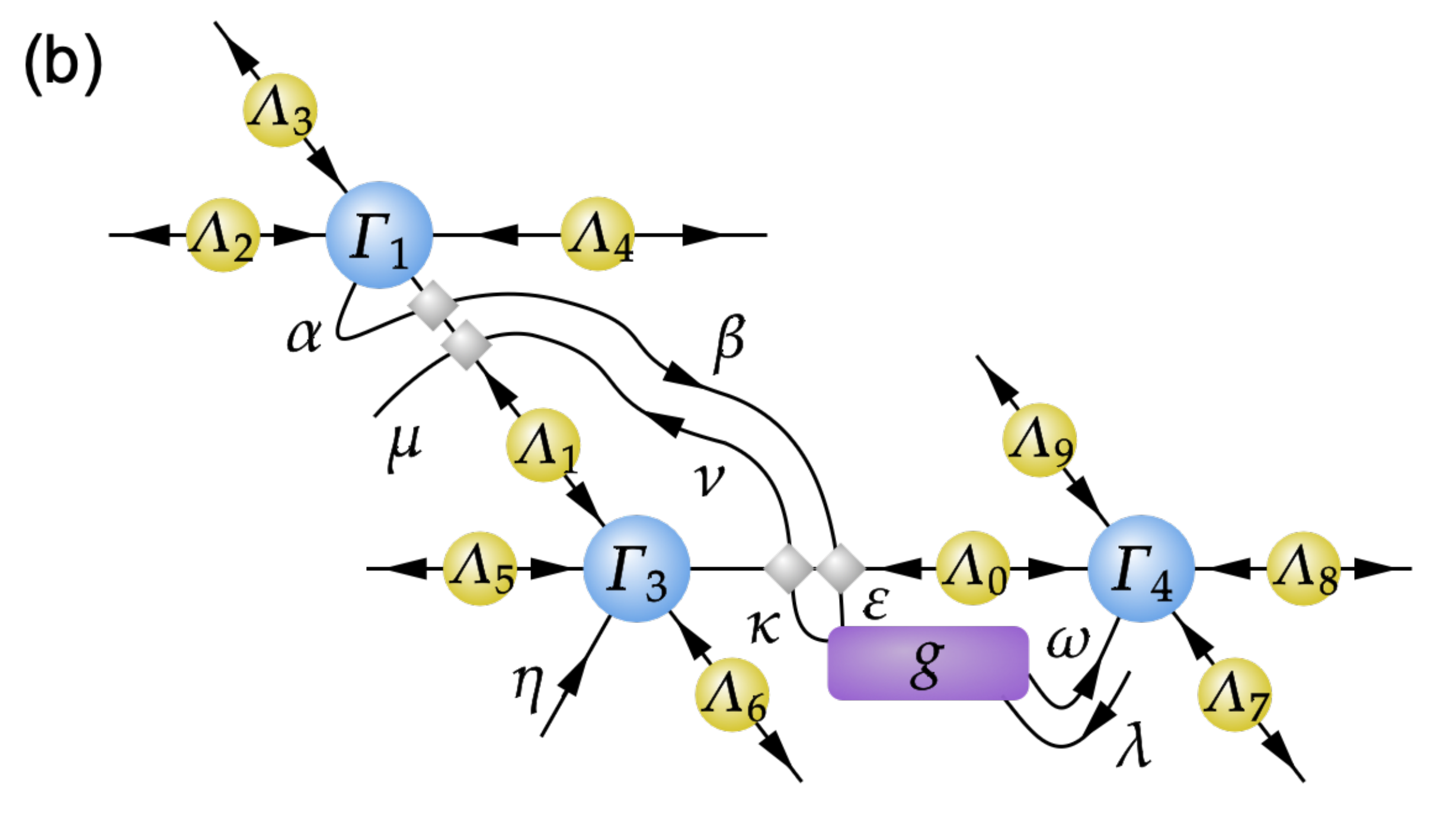}\\[1em]
    \else\hspace*{-6em}\includegraphics[scale=0.16]{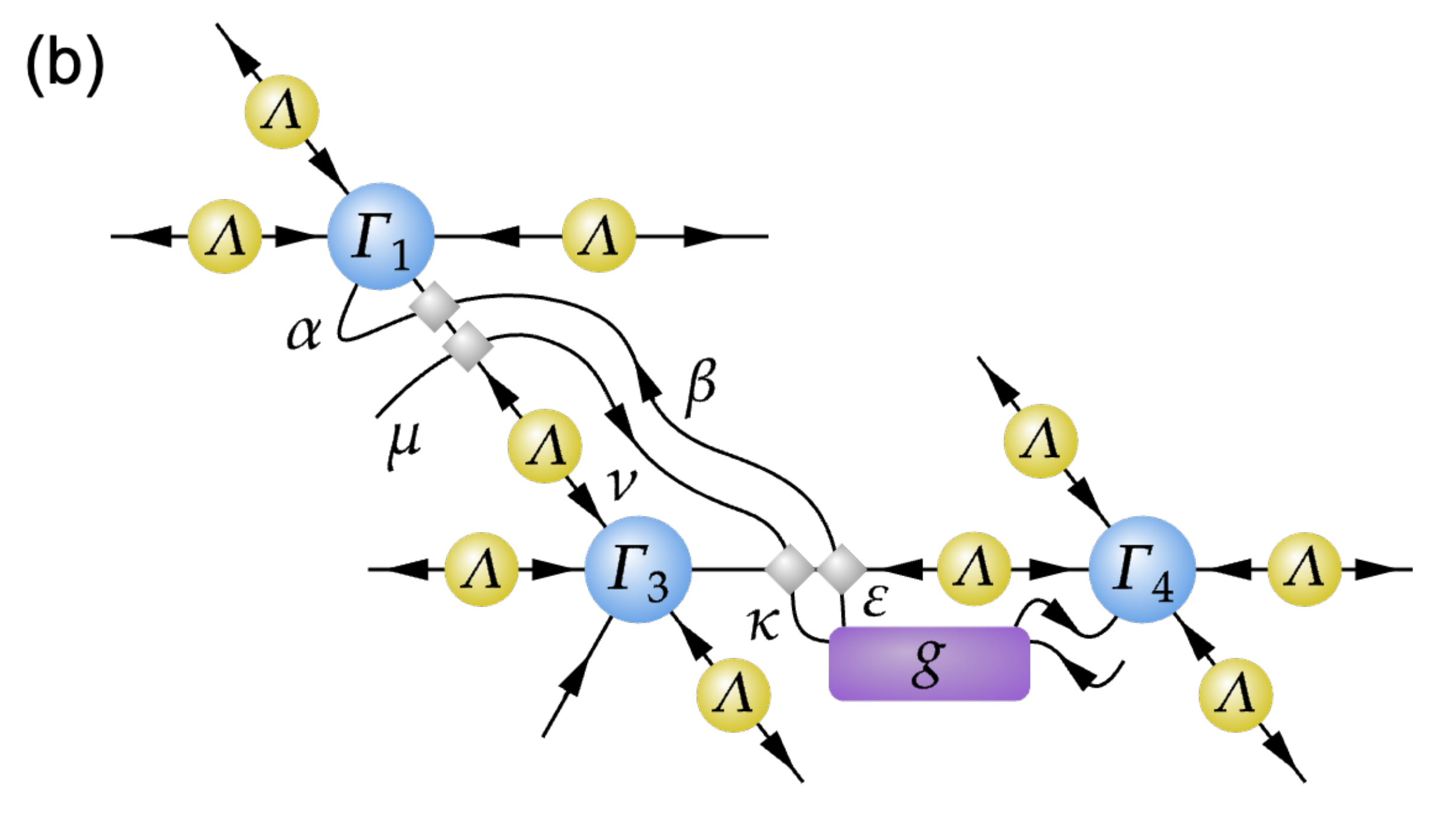}\\[1em]\fi
    \hspace*{8em}\includegraphics[scale=0.16]{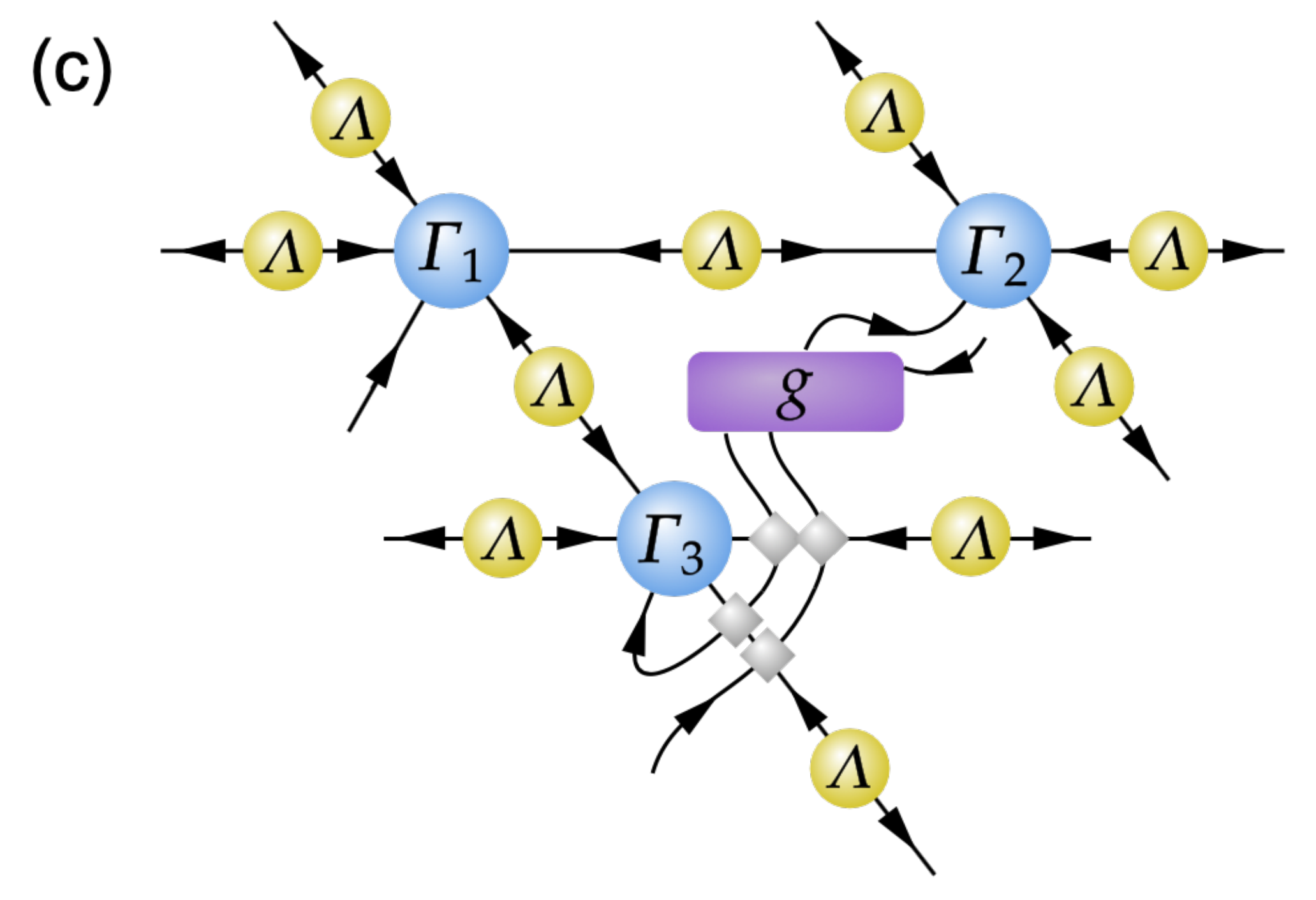}\\[-7.4em]
    \hspace*{-10em}\includegraphics[scale=0.16]{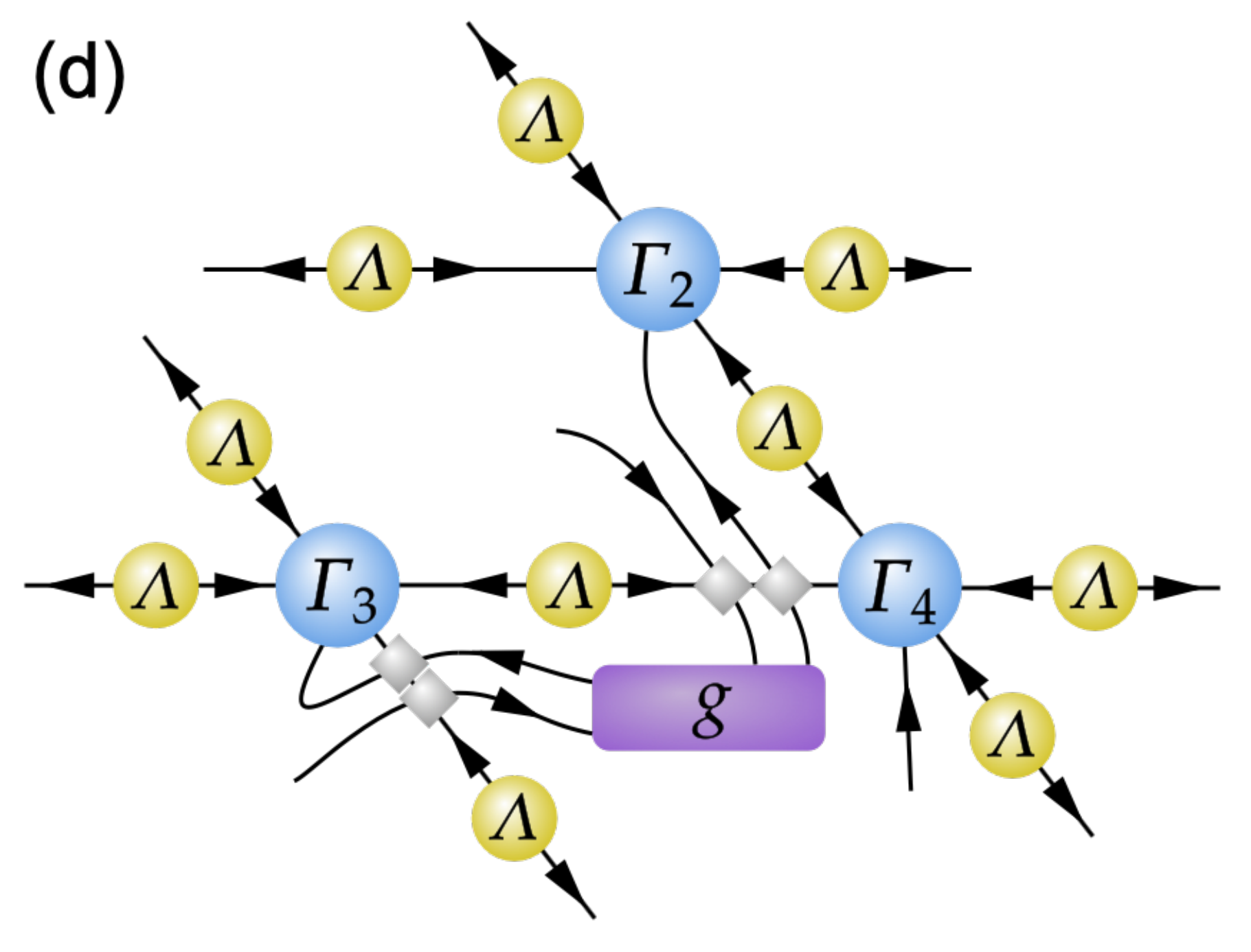}
    \begin{minipage}{0.45\textwidth}
    \vspace{0.6em}
    \caption{Tensor network diagrams for the refined simple update scheme for NNN terms. Four diagrams are needed to update the relevant tensors. We use the standard $\varGamma$-$\varLambda$ form of iPEPS tensor network \cite{Corboz2010-PEPS-NN}. $g$ is the Trotter gate for the corresponding NNN term. Arrows indicate the flow of quantum numbers \cite{Weichselbaum2012-QSpace}. Grey rhombuses depict the fermionic swap gates \cite{Corboz2010-PEPS-NN}.} 
    \vspace{-0.4em}
    \label{SUDiagram}
    \end{minipage}
\end{figure}

In our research, we implement a refined version of simple update based on \cite{Corboz2010-PEPS-NNN}, which leads to tremendous speedup of the algorithm. The update of the nearest neighbor (NN) terms (including on-site) is the same as described in \cite{Corboz2010-PEPS-NN}. We modified the NNN update scheme, in a manner suggested by Andreas Weichselbaum as shown in Fig.~\ref{SUDiagram}, for further processing.

\begin{figure}[htp!]
    \vspace{1em}
    \centering
    \ifdefined\Publication
    \includegraphics[scale=0.16]{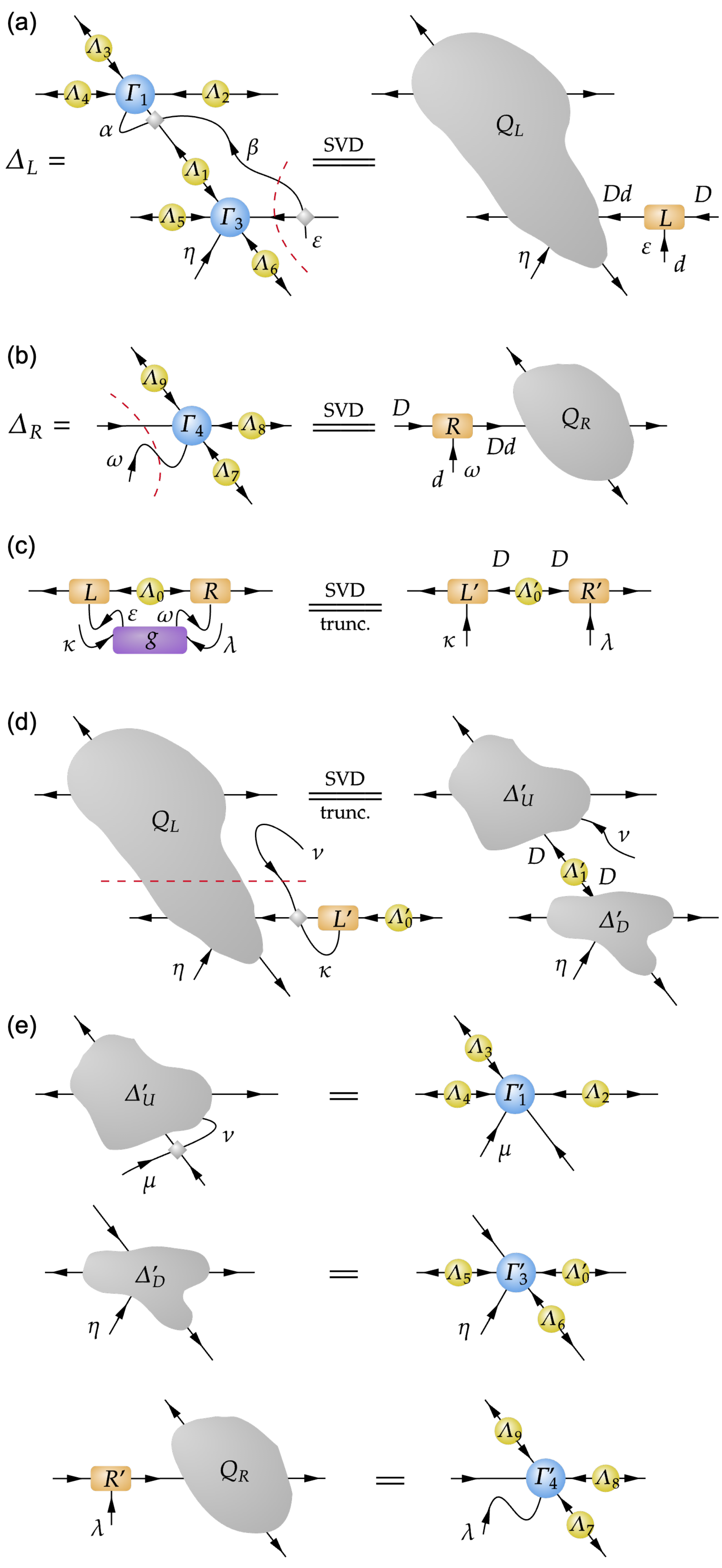}
    \else \includegraphics[scale=0.16]{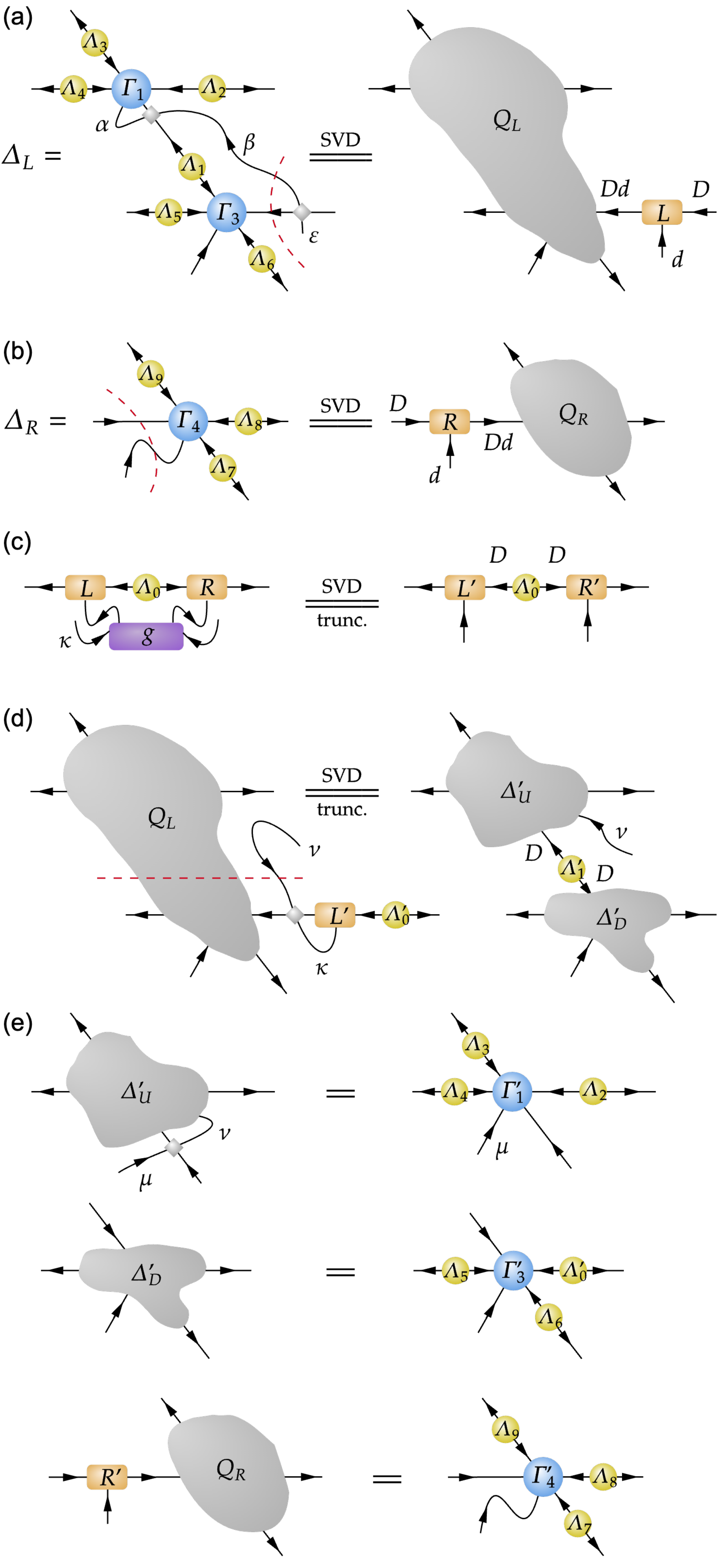} \fi
    \begin{minipage}{0.45\textwidth}
    \vspace{4pt}
    \caption{Realization of the diagram in Fig.~\ref{SUDiagram}(b), which updates $\varLambda_0$, $\varLambda_1$ and all $\varGamma$s, while leaving all other $\varLambda$s unchanged. This process \emph{factors out} the environment encoded in $Q_L$ and $Q_R$, before applying the Trotter gate. (a)(b) Construct tensors $\varDelta_L$ and $\varDelta_R$ and split off the relevant physical and auxiliary indices using SVD to obtain tensors $L$ and $R$. (c) Apply the Trotter gate to tensor $L$, $R$ and the corresponding $\varLambda$. Obtain the updated $L'$, $R'$ and $\varLambda'$ via truncated SVD. (d) Restore the $\varGamma$-$\varLambda$ structure via truncated SVD. (e) Update all the involved $\varGamma$ and $\varLambda$ tensors. Greek labels, in correspondence with Fig.~\ref{SUDiagram}(b), keep track of the physical indices.}
    \label{SUDetails}
    \end{minipage}
\end{figure}

The major improvement comes from the separation of indices directly involved in the update from the other \emph{spectator} indices not involved in the update. Figure \ref{SUDetails} shows the details for treating the diagram in Fig.~\ref{SUDiagram}(b). The other diagrams can be performed analogously. Greek letters $\alpha,\beta,\ldots$ label the physical indices. Figure \ref{SUDetails}(a) describes the construction of tensor $\varDelta_L$ and the application of corresponding swap gates. Then, we split off the relevant physical and auxiliary indices via singular value decomposition (SVD). Next, we perform analogous actions for $\varDelta_R$ as depicted in Fig.~\ref{SUDetails}(b). These procedures generate tensors $L$ and $R$ of size $Dd×D×d$, where $D$ is the bond dimension of the auxiliary index and $d$ the dimension of local Hilbert space. Hereafter, we apply the Trotter gate, as in Fig.~\ref{SUDetails}(c), and obtain the updated tensors $L'$ and $R'$ via truncated SVD. Then, we restore the original $\varGamma$-$\varLambda$ structure via truncated SVD as shown in Fig.~\ref{SUDetails}(d). Finally, we update all the involved $\varGamma$ and $\varLambda$ tensors as described in Fig.~\ref{SUDetails}(e). The last two steps recover the original structure in Fig.~\ref{SUDiagram}(b).

The refined simple update scheme has a much better scaling behavior with respect to the bond dimension $D$. The original version of NNN simple update \cite{Corboz2010-PEPS-NNN} involves an SVD of complexity $O(D^{11}d^4)$, which is also the dominant contribution to the total complexity of the algorithm. In our refined scheme, the dominant complexity comes from the SVD in Fig.~\ref{SUDetails}(a), which scales as $O(D^7d^3)$, much lower than the complexity of the original simple update scheme.

\begin{figure}[h!]
    \vspace*{0.7em}
    \centering
    \hspace{-15pt}
    \ifdefined\Publication
    \includegraphics[width=0.46\textwidth]{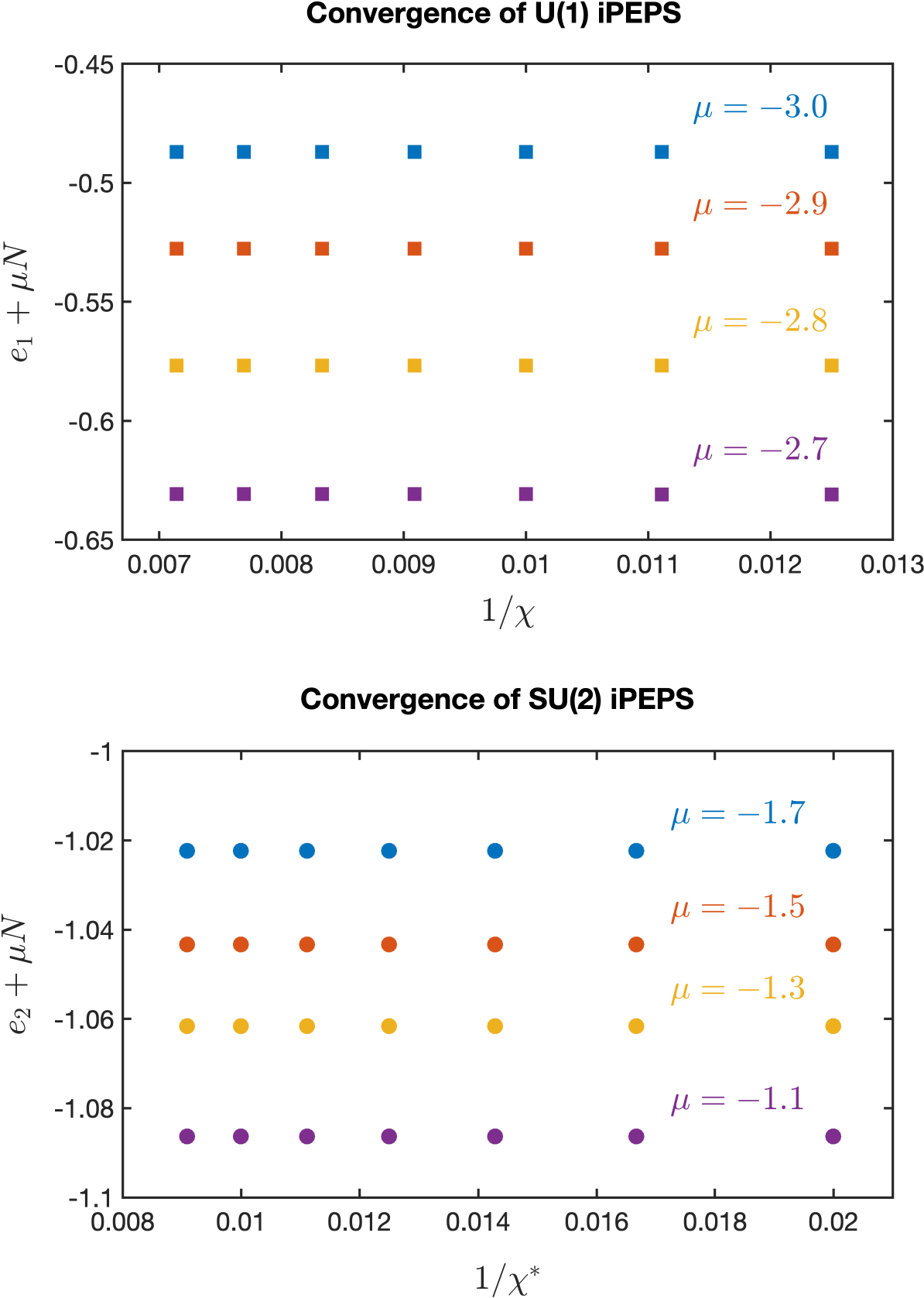}
    \else\includegraphics[width=0.46\textwidth]{../images/Convergence.png}\\ \fi
    \begin{minipage}{0.45\textwidth}
    \caption{The convergence characteristics of the $\mathrm{U}(1)$ and $\mathrm{SU}(2)$ iPEPS with respect to the environmental bond dimension $\chi$ and $\chi^*$. The energies $\eA-\mu N$ and $\eS-\mu N$ have been well converged starting from $\chi\geqslant80$ and $\chi^*\geqslant50$.\vspace{-1pt}}
    \label{Convergence}
    \end{minipage}
\end{figure}

\begin{figure*}[htp!]
    \centering
    \vspace*{0.3em}
    \includegraphics[width=0.95\textwidth]{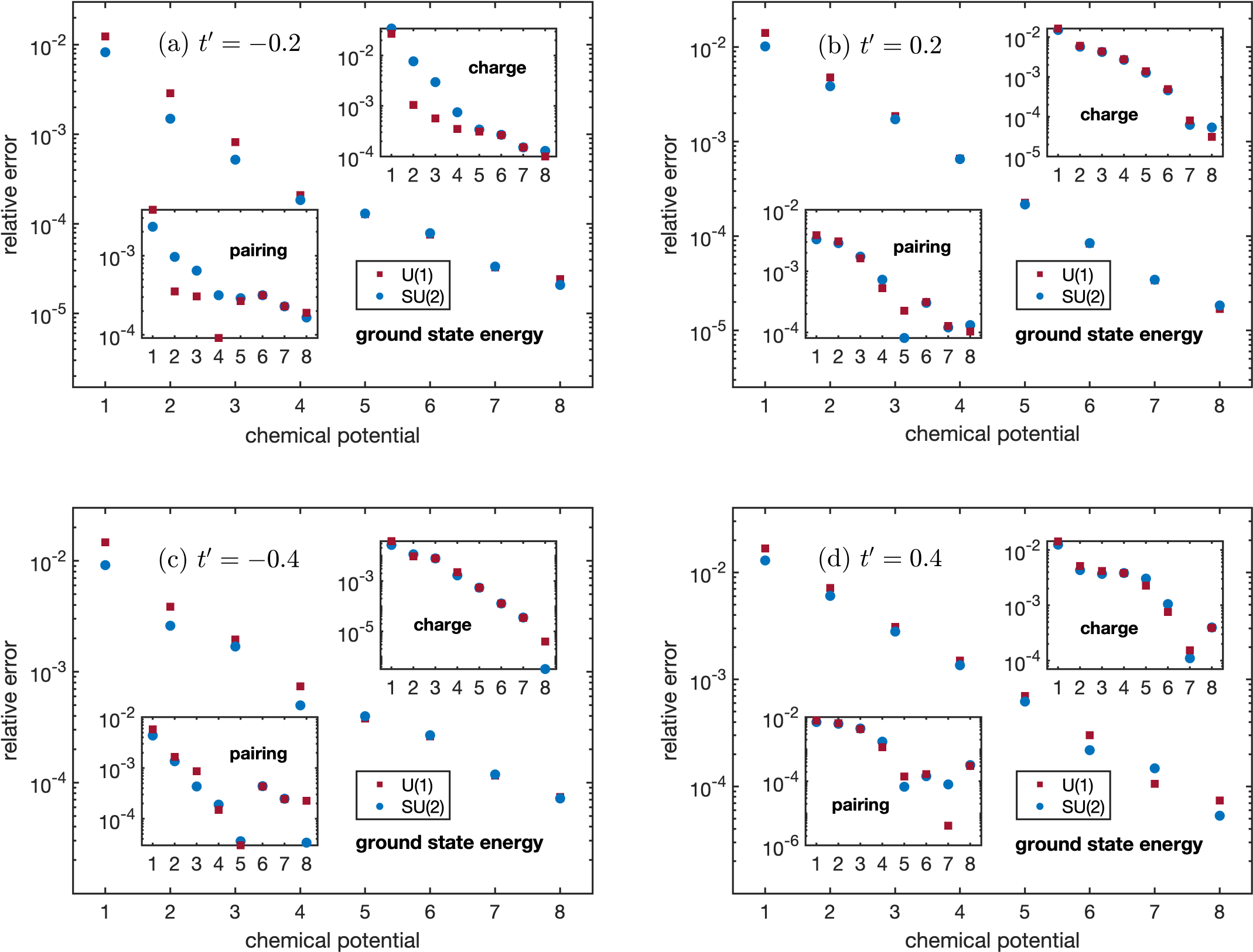}
    \begin{minipage}{0.9\textwidth}
    \vspace*{0.7em}
    \caption{The error of ground state energy $e_0$, charge density $n_0$ and singlet pairing amplitude $\Delta_s$ of the free-fermion model obtained by $\mathrm{U}(1)$ and $\mathrm{SU}(2)$ iPEPS algorithm on a 2×2 supercell at $D=9$ and $D^*[D] = 6[9]$, respectively. We set $\gamma=1$ and focus on four values of $\tnnn\! = \pm0.2, \pm0.4$. At each $\tnnn$, we vary chemical potential from $\mu=1$ to $8$. The accuracy at large chemical potential is generally better than that at small chemical potential, which is consistent with the fact that the energy gap grows with the chemical potential in this model.}
    \label{Benchmark}
    \end{minipage}
\end{figure*}

\vspace{-2pt}
\section{Technical Details of the $\mathrm{SU}(2)$ iPEPS Simulations}
\label{Technical}
\vspace{-2pt}

Our computations start with a random initial iPEPS tensor network state written in the $\varGamma$-$\varLambda$ form \cite{Corboz2010-PEPS-NN}. The ground state is attained by repeatedly applying the projector $\exp\{-\tau(\mathcal{H}+\mu N)\}$ ($\tau$ is a small number, $\mathcal{H}$ the Hamiltonian \eqref{Hamiltonian}, $\mu$ the chemical potential and $N$ the charge density) to the initial state. The application is broken into a sequence of projectors involving only NN or NNN terms via Suzuki-Trotter decomposition. The Trotter error is of order $O(\tau^2)$. Therefore, we start with a large $\tau$ in the beginning so as to approach the ground state quickly, followed by a gradual decrease of $\tau$ to improve the accuracy.

Empirically, we find $\tau=0.1$ a fairly good starting point. During the update iteration, we measure the expectation value of the Trotter gate and compute an estimation of the ground state energy, which serves as an indicator of convergence. $\tau$ is reduced by half once the decrease of estimated ground state energy in an update step drops below $\tau^2$. We regard the convergence to be reached when $\tau$ drops below the threshold of $10^{-3}$. 

The optimization starts with $D=2$ for $\mathrm{U}(1)$ iPEPS and $D^*=2$ for $\mathrm{SU}(2)$ iPEPS. After reaching convergence at the fixed bond dimension, we increment $D$ or $D^*$ by $1$ (but see also Sec.~\ref{Guidance}), until we arrive at $D\!=\!12$ for $\mathrm{U}(1)$ or $D^*\!=\!7$ for $\mathrm{SU}(2)$. A preliminary ground state is obtained for some specific chemical potential $\mu_0$, which serves as the initialization of the optimization for all other values of $\mu$.

The symmetry bookkeeping of the tensors 
in our algorithm is managed via the QSpace tensor library \cite{Weichselbaum2012-QSpace,Weichselbaum2012-QSpace-XSymbols,osqspacev4}. The QSpace library implements a series of tensor operations (e.g. contraction, SVD, etc.) that handle the propagation of symmetries. Therefore, we are able to use identical codes for both our $\mathrm{U}(1)$ and $\mathrm{SU}(2)$ iPEPS, differing only in the initialization where symmetries are designated. This guarantees the equal reliability of our iPEPS implementation with different symmetries.

The accurate contraction and measurement of the observables are performed via the Corner Transfer Matrix Renormalization Group (CTMRG) scheme \cite{Corboz&Vidal2011-tJ-PEPS&DMRG,Corboz2014-tJ-PEPS,Bruognolo2020-iPEPS-Review}, which generates a series of environmental tensors. The convergence of the ground state energies $\eA+\mu N$ and $\eS+\mu N$ with respect to environmental bond dimension $\chi$ for $\mathrm{U}(1)$ iPEPS or $\chi^*$ for $\mathrm{SU}(2)$ iPEPS is shown in Fig.~\ref{Convergence}. We find that they already reach sufficient convergence at $\chi=80$ and $\chi^*=50$. For safety, we set $\chi$ is set to be $\chi=144$ and $\chi^*=100$ in the main text.

\section{Benchmark}
\label{BenchmarkSection}

In this section, we present benchmarks of our $\mathrm{U}(1)$ and $\mathrm{SU}(2)$ iPEPS algorithms for an exactly solvable free-fermion model. The Hamiltonian reads:
\begin{equation}
\label{FFSModel}
\begin{split} 
    \mathcal{H} &= -\tnn \sum_{\langle i,j\rangle,\sigma} \left[ c^\dagger_{i\sigma} c_{j\sigma} + \gamma \Delta_{ij} + \text{h.c.}\right] \\
    &\;\; - \tnnn\!\! \sum_{\langle\!\langle i,j\rangle\!\rangle,\sigma} \left[ c^\dagger_{i\sigma} c_{j\sigma} + \gamma \Delta_{ij} + \text{h.c.}\right] + \mu \sum_{i,\sigma} c^\dagger_{i\sigma} c_{i\sigma},
\end{split}
\end{equation}
where \vspace{-2pt}
\begin{equation}
    \Delta_{ij} = c_{i\uparrow} c_{j\downarrow} - c_{i\downarrow} c_{j\uparrow},
\end{equation}
$\tnn$ and $\tnnn$ are the nearest and next-nearest neighbor hopping amplitude, $\gamma$ the singlet pairing potential, and $\mu$ the chemical potential. The ground state energy can be computed via a Fourier transform followed by a Bogoliubov transform:
\begin{equation}
\vspace{2.5pt}
\begin{aligned}
    e_0 &= \sum_\mathbf{k} \left[\xi_\mathbf{k} - \sqrt{\xi_\mathbf{k}^2 + \Delta_\mathbf{k}^2}\right],\\
    \xi_\mathbf{k} &= \mu - 2\tnn(\cos\mathbf{k}\!\cdot\!\mathbbm{1}_\mathrm{x} + \cos\mathbf{k}\!\cdot\!\mathbbm{1}_\mathrm{y})\\ &\qquad - 2 \tnnn(\cos\mathbf{k}\!\cdot\!\mathbbm{1}_\mathrm{d} + \cos\mathbf{k}\!\cdot\!\mathbbm{1}_\mathrm{o}), \\
    \Delta_\mathbf{k} &= 2\gamma\tnn(\cos\mathbf{k}\!\cdot\!\mathbbm{1}_\mathrm{x} + \cos\mathbf{k}\!\cdot\!\mathbbm{1}_\mathrm{y})\\ &\: + 2 \gamma \tnnn(\cos\mathbf{k}\!\cdot\!\mathbbm{1}_\mathrm{d} + \cos\mathbf{k}\!\cdot\!\mathbbm{1}_\mathrm{o}),
\end{aligned}
\end{equation}
and $\mathbbm{1}_\mathrm{x}$, $\mathbbm{1}_\mathrm{y}$, $\mathbbm{1}_\mathrm{d}$, $\mathbbm{1}_\mathrm{o}$ are the unit vectors of horizontal, vertical, diagonal and off-diagonal directions, respectively. In all our benchmarks, we set $t=1$ and $\gamma=1$. And the analytical numerics are performed on a 100×100 finite size lattice, which suffices to approximate the thermodynamic limit with a double precision accuracy.

\begin{figure}[t!]
    \vspace{1em}
    \centering
    \ifdefined\Publication \vspace{-1.7em}
    \includegraphics[width=0.46\textwidth]{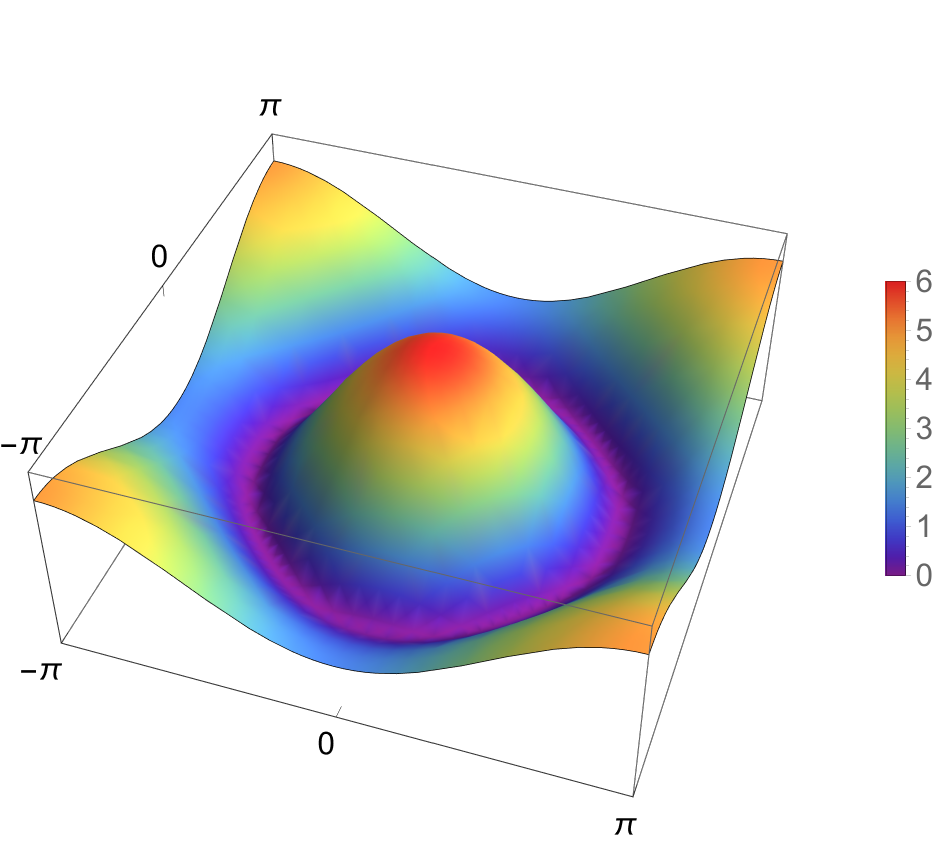}\vspace{1.7em}
    \else\includegraphics[width=0.46\textwidth]{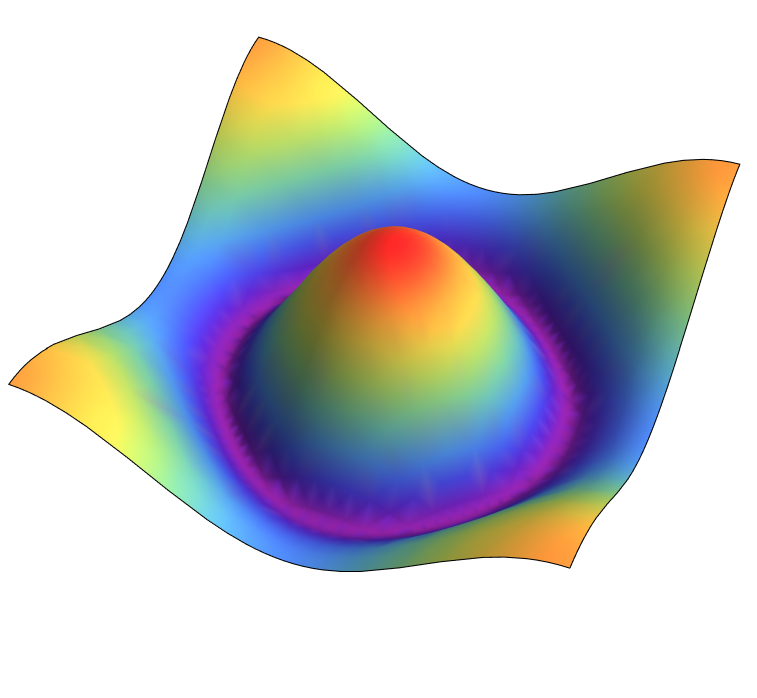}\\ \fi
    \begin{minipage}{0.45\textwidth}
    \vspace{-10pt}
    \caption{The dispersion relation for the free-fermion model defined according to Eq.~(\ref{FFSModel}) at $t=1$, $t'=0.2$, $\gamma=1$, and $\mu=1$. This system demonstrates a gapless nature when $\mu=0$, and develops an energy gap for $\mu>0$. The dispersion relation exhibits a characteristic \emph{Mexican hat} shape for $0<\mu\leqslant8$. In this range, the energy gap is quantified by the minimal value around the depicted purple region.}
    \label{Dispersion}
    \end{minipage}
\end{figure}

In the case of $\gamma=1$, the dispersion relation $E_\mathbf{k} = \sqrt{\xi_\mathbf{k}^2 + \Delta_\mathbf{k}^2}$ has a simpler form 
\begin{equation}
    E_\mathbf{k} = \sqrt{(\mu - \Delta_\mathbf{k})^2 + \Delta_\mathbf{k}^2},
\end{equation}
which exhibits a \emph{Mexican hat} shape, as depicted in Fig.~\ref{Dispersion}. The minimum of the dispersion relation satisfies $\partial_\mathbf{k} E_\mathbf{k} = 0$. For $0\!<\!\mu\!\leqslant\!8$, this condition leads to $\mu = 2\Delta_\mathbf{k}$, which yields an energy gap
\begin{equation}
    \label{egap}
    \Delta_E = \mu / \sqrt{2}
\end{equation}
Therefore, the energy gap grows as the chemical potential increases. Also, we can calculate analytically the average charge density $n_0$ and singlet pairing amplitude $\Delta_s$:
\begin{equation}
\begin{aligned}
    n_0 &= \sum_{\mathbf{k}}\left[ 1 - \frac{\xi_\mathbf{k}}{\sqrt{\xi_\mathbf{k}^2 + \Delta_\mathbf{k}^2}}\right], \\
    \Delta_s &= \sum_{\mathbf{k}} \Delta^\mathrm{nn}_\mathbf{k} \cdot \frac{\Delta_\mathbf{k}}{2\sqrt{\xi_\mathbf{k}^2 + \Delta_\mathbf{k}^2}},
\end{aligned}
\end{equation}
where
\begin{equation}
    \Delta^\mathrm{nn}_\mathbf{k} = 2\gamma\tnn(\cos\mathbf{k}\!\cdot\!\mathbbm{1}_\mathrm{x} + \cos\mathbf{k}\!\cdot\!\mathbbm{1}_\mathrm{y})
\end{equation}

Benchmark results for the $\mathrm{U}(1)$ and $\mathrm{SU}(2)$ iPEPS algorithms are depicted in Fig.~\ref{Benchmark}, demonstrating their errors relative to the aforementioned analytical values on a 2×2 supercell with $D=9$ and $D^*[D] = 6[9]$, respectively. The analysis incorporates four selected values of $\tnnn = \pm0.2, \pm0.4$, and for each $\tnnn$ value, the chemical potential is varied from $\mu=1$ to $8$.

As the chemical potential increases, there is a discernible enhancement in the accuracy of both the $\mathrm{U}(1)$ and $\mathrm{SU}(2)$ iPEPS algorithms. This improvement can be attributed to the increase of the energy gap in proportion to the chemical potential, as described by Eq.~(\ref{egap}). An increased energy gap implies diminished long-range correlations, thereby enabling the ground state properties to be better encoded within an iPEPS ansatz at a limited bond dimension.

The $\mathrm{SU}(2)$ symmetry is an inherent characteristic of the ground state of this free-fermion model. Thus, the errors produced by both the $\mathrm{U}(1)$ and $\mathrm{SU}(2)$ iPEPS are comparable. However, in certain instances, the $\mathrm{SU}(2)$ iPEPS demonstrates a marginally superior accuracy, which can likely be attributed to the additional $\mathrm{SU}(2)$ symmetry constraint that aids in better approximating the true $\mathrm{SU}(2)$ symmetric ground state. For large chemical potential, the accuracies of both algorithms reach $10^{-4}$, corroborating the efficacy of the algorithms.

\section{Guided and Unguided $\mathrm{U}(1)$ iPEPS Simulations}
\label{Guidance}

As a mathematical construction, the $\mathrm{U}(1)$ iPEPS ansatz encompasses a larger parameter space compared to the $\mathrm{SU}(2)$ iPEPS ansatz at the same virtual bond dimensions, since the former is subject to fewer symmetry constraints. Consequently, it might be simplistically perceived that the $\mathrm{U}(1)$ iPEPS should be capable of capturing $\mathrm{SU}(2)$ symmetric states for the parameter regime where the ground states retain $\mathrm{SU}(2)$ symmetry. Yet, from a pragmatic perspective, the $\mathrm{U}(1)$ iPEPS rarely converges to an $\mathrm{SU}(2)$ symmetric state. This section aims to provide theoretical insights into this artifact.

\vspace{1em}
\begin{table}[ht]
\renewcommand{\arraystretch}{1.2}
\rule{.47\linewidth}{1.2pt}\\[2pt]
\begin{tabular}{ccc}
$D^*$ & $D$ & $S[R]$\\[2pt]
\hline\\[-10pt]
1 & 1 & $0[1]$\\
2 & 3 & $0[1]\oplus\frac{1}{2}[1]$\\
3 & 4 & $0[2]\oplus\frac{1}{2}[1]$\\
4 & 6 & $0[2]\oplus\frac{1}{2}[2]$\\
5 & 7 & $0[3]\oplus\frac{1}{2}[2]$\\
6 & 9 & $0[3]\oplus\frac{1}{2}[3]$\\
7 & 12 & $\quad0[3]\oplus\frac{1}{2}[3]\oplus1[1]$\\
8 & 13 & $\quad0[4]\oplus\frac{1}{2}[3]\oplus1[1]$\\
\vdots & \vdots & \vdots\\
\end{tabular}\\
\hspace{2pt}\rule{.47\linewidth}{1.2pt}
\begin{minipage}{0.45\textwidth}
\caption{The $D^*$-$D$ correspondence and the details of the symmetry sectors. $S$ is the total spin quantum number and $R$ the number of spin-$S$ multiplets, each of dimension $2S+1$. For example, for $D^*=5$, there are 5 multiplets, namely 3 singlets and 2 doublets, leading to a total bond dimension of $D=7$ ($=3\times1+2\times2$).}
\label{DDTable}
\end{minipage}
\end{table}

During the optimization of tensor networks, specific strategies are usually implemented to foster a more efficient approximation of the ground state. Concretely, the usual approach commences with a small bond dimension $D$, e.g. $D\!=\!2$, followed by optimization to convergence, prior to incrementing $D$ until the pre-determined maximum $D$ is reached. The primary rationale behind this strategy is the circumvention of potential local minimums in the expansive parameter space spanned by a large $D$. Nevertheless, this strategy engenders complications when searching through the $\mathrm{SU}(2)$ symmetric subspace.

In the course of conducting our $\mathrm{SU}(2)$ iPEPS simulations, we have identified a distinct relationship between the number of kept multiplets $D^*$ and the corresponding bond dimension $D$. Particularly, in the context of the $\tnn$-$\tnnn$ Hubbard model, the $D^*$-$D$ correspondence and the details of the symmetry sectors are delineated in Table \ref{DDTable}.

A perusal of the table reveals that not all choices of $D$ are congruous with the $\mathrm{SU}(2)$ symmetry. In other words, if one carries out an optimization at $D_u\!=\!2,5,8,10,11,\cdots$ (the $D$ values not in Table~\ref{DDTable}), the resulting ground state is invariably destined to violate the $\mathrm{SU}(2)$ symmetry. Hence, should one want to preserve of $\mathrm{SU}(2)$ symmetry within a $\mathrm{U}(1)$ iPEPS algorithm, the optimization process should be conducted exclusively following the sequence $D_g\!=\!1,3,4,6,7,9,12,\cdots$. We refer to this approach as \emph{guided} $\mathrm{U}(1)$ optimization. Conversely, should one traverse the entire sequence $D\!=\!1,2,3,4,5,6,7,8,9,10,11,12,\cdots$, the $\mathrm{SU}(2)$ symmetry is assured to be broken. We thus refer to this method as \emph{unguided} $\mathrm{U}(1)$ optimization.

\begin{figure}[htp!]
    \vspace*{1em}
    \centering
    \hspace{-15pt}
    \ifdefined\Publication
    \includegraphics[width=0.46\textwidth]{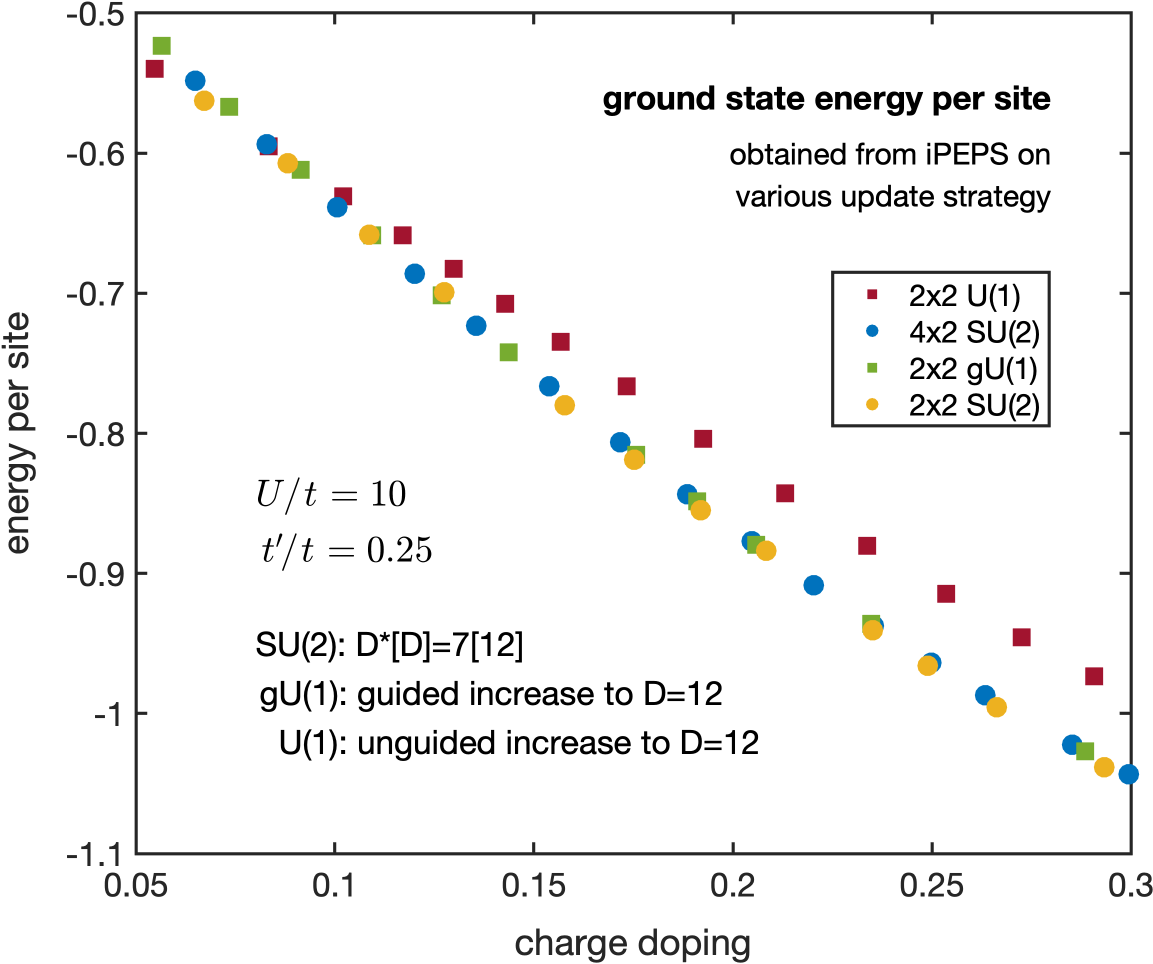}\\
    \else\includegraphics[width=0.46\textwidth]{img/GE22_SE.pdf}\\ \fi
    \begin{minipage}{0.45\textwidth}
    \caption{Ground state energy per site obtained from guided and unguided $\mathrm{U}(1)$ iPEPS on the 2×2 supercells, as well as $\mathrm{SU}(2)$ iPEPS on the 4×2 and 2×2 supercells respectively as a function of doping. The data points of 2×2 guided $\mathrm{U}(1)$, 2×2 $\mathrm{SU}(2)$ and 4×2 $\mathrm{SU}(2)$ computations overlap with each other, confirming the uniformity of the $\mathrm{SU}(2)$ ground states.}
    \label{GE22}
    \end{minipage}

    \vspace{20pt}
    \centering
    \hspace{-15pt}
    \ifdefined\Publication
    \includegraphics[width=0.4\textwidth]{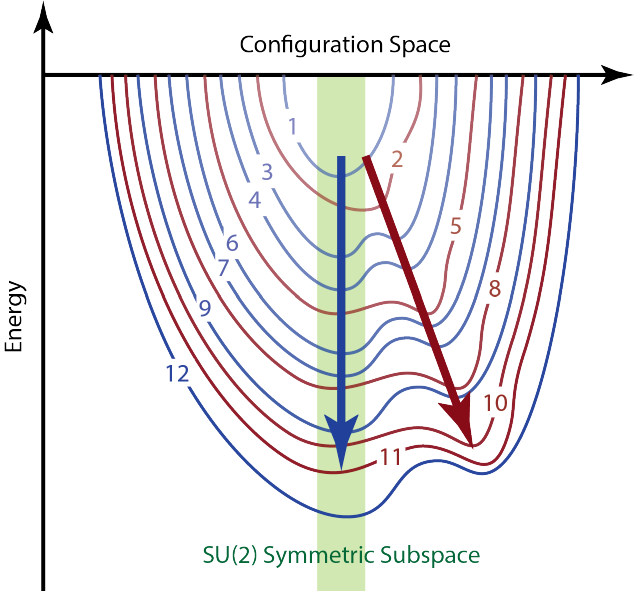}\\
    \else\includegraphics[width=0.4\textwidth]{img/ConfigSketch.pdf}\\ \fi
    \begin{minipage}{0.45\textwidth}
    \vspace{7pt}
    \caption{Schematic depiction of the energy landscape of the $\mathrm{U}(1)$ iPEPS configuration space for the $\tnn$-$\tnnn$ Hubbard model, specifically within parameter settings where ground states preserve $\mathrm{SU}(2)$ symmetry. Each curve depicted in the diagram represents energy at varying points in the configuration space, with the numbers indicating the corresponding bond dimension. The blue (red) curves correspond to bond dimensions compatible (incompatible) with the $\mathrm{SU}(2)$ symmetry. When optimizing $\mathrm{U}(1)$ iPEPS tensor networks, a guided optimization process would follow the direction indicated by the blue arrow, ultimately reaching the $\mathrm{SU}(2)$ symmetric ground state. Conversely, an unguided optimization process would follow the direction of the red arrow, leading to entrapment at a local minimum.}
    \label{Configs}
    \end{minipage}
\end{figure}

Figure \ref{GE22} shows the ground state energy per site obtained from both guided and unguided $\mathrm{U}(1)$ iPEPS on the 2×2 supercells, in conjunction with the $\mathrm{SU}(2)$ iPEPS on the 4×2 and 2×2 supercells. Remarkably, the 2×2 guided $\mathrm{U}(1)$, the 2×2 $\mathrm{SU}(2)$, and the 4×2 $\mathrm{SU}(2)$ results agree very well with each other. The overlap between 2×2 and 4×2 $\mathrm{SU}(2)$ iPEPS substantiates the uniformity of the $\mathrm{SU}(2)$ ground states. Furthermore, the alignment between 2×2 guided $\mathrm{U}(1)$ and 2×2 $\mathrm{SU}(2)$ iPEPS signifies that they are producing states that are physically analogous. This proposition is further supported by comparisons of other observables.

Conversely, the unguided $\mathrm{U}(1)$ iPEPS yields non-$\mathrm{SU}(2)$ symmetric states characterized by local magnetic orders, which exhibit a higher energy than the states produced by the other three methodologies. This suggests that an iPEPS preserving only the $\mathrm{U}(1)$ spin symmetry can become entrenched in specific local minimums in the absence of appropriate guidance or additional symmetry constraints.


An intuitive interpretation of this phenomenon is rooted in the observation that the configuration space of the $\mathrm{U}(1)$ iPEPS for the $\tnn$-$\tnnn$ Hubbard model is non-convex. This is depicted in Fig.~\ref{Configs}, a sketch of the energy landscape of such a configuration space. In this particular instance, two local minimums are present (though there could potentially be more, we limit our focus to the two most relevant local minimums): one resides within the $\mathrm{SU}(2)$ symmetric subspace; the other stays outside. It becomes evident that when the algorithm adheres to the guided sequence $D_g$, it remains within the $\mathrm{SU}(2)$ symmetric subspace (blue arrow). However, in the absence of such a guidance, the algorithm veers away from the $\mathrm{SU}(2)$ subspace, beginning from $D=2$ (red arrow). Hence, we have demonstrated that guided and unguided $\mathrm{U}(1)$ iPEPS can lead to markedly different results.

\newpage

\section{Singlet Pairing Properties of the $\mathrm{SU}(2)$ Ground States}
\label{Singlet}

\begin{figure}[htp!]
    \vspace{8pt}
    \centering
    \ifdefined\Publication
    \includegraphics[width=0.36\textwidth]{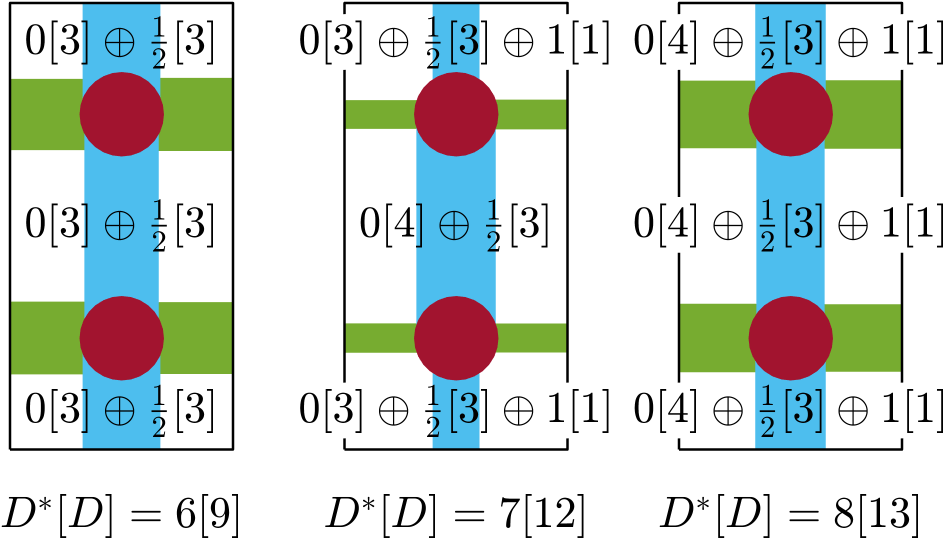}\\
    \else\includegraphics[width=0.36\textwidth]{img/PairingD.pdf}\\ \fi
    \begin{minipage}{0.45\textwidth}
    \caption{Singlet pairing properties of the $\mathrm{SU}(2)$ ground states at different bond dimensions with $\delta\approx0.3$ and $\tnnns/\tnn=0.25$. A 1×2 subcell of each 4×2 supercell is displayed (since the states are uniform in the horizontal direction). At $D^*[D]\!=\!6[9]$ and $D^*[D]\!=\!8[13]$, all horizontal and vertical bonds host the same symmetry sectors. By contrast, at $D^*[D]\!=\!7[12]$, the central vertical bond has one more singlet and one less triplet compared to the other bonds. This results in an enhanced singlet pairing amplitude in the central vertical bonds. Increasing $D^*$ to $D^*[D]\!=\!8[13]$ rectifies this anomaly.} 
    \label{PairingD}
    \end{minipage}
    \vspace*{2pt}
\end{figure}

Our $\mathrm{SU}(2)$ iPEPS simulations encounter a minor artifact at $D^*[D]\!=\!7[12]$ when measuring the singlet pairing amplitudes. Figure~\ref{PairingD} displays a 1×2 subcell as a representative of each 4×2 supercell, since the states are uniform in the horizontal direction. For the particular setting of $D^*[D]\!=\!7[12]$, the vertical bond in the middle exhibits a different symmetry structure \ifdefined\Publication than \else compared with \fi the other bonds. This discrepancy significantly impacts the measured singlet pairing properties, as they deviate markedly from the expected $d$-wave characteristics.

We have confirmed that this anomaly arises exclusively at $D^*[D]\!=\!7[12]$. At adjacent $D^*[D]\!=\!6[9]$ and $D^*[D]\!=\!8[13]$ (as well as all smaller $D^*$), the symmetry structure remains consistent across all vertical and horizontal bonds, and the singlet pairing amplitudes demonstrate excellent $d$-wave characteristics, as illustrated in Fig.~\ref{PairingD}. Thus, we conclude that the ground state should exhibit a $d$-wave pairing order. In the main text, we present the pairing properties as derived from $D^*[D]\!=\!8[13]$.


\section{Stripes with Longer Periods}
\label{LongStripes}

In the main text, our focus is predominantly on the period 4 stripe states. Nevertheless, it is important to note that stripes with longer periods might emerge under varying parameter configurations. This section is dedicated to a further exploration of this aspect. As the manifestation of stripes with longer periods necessitates iPEPS on larger supercells, the computational cost can become significantly higher. Consequently, our objective here is not a complete optimization of the iPEPS across all scanning points. Rather, our aim is to estimate the region where the ground state begins to deviate from the charge uniform state.

\ifdefined\Publication We initiate \else Our approach entails initiating \fi the iPEPS simulation with a uniform state, obtained through guided $\mathrm{U}(1)$ iPEPS at large doping. Consider, for instance, a period 8 stripe. A 16×2 iPEPS is necessary to capture stripes characterized by period 8 charge orders. We acquire a single ground state via the guided 2×2 $\mathrm{U}(1)$ iPEPS at large doping, typically around $\delta\!\approx\!0.3$. Its 2×2 supercell is then replicated 8 times to construct a 16×2 supercell. The corresponding ground state subsequently serves as the initialization for the further 16×2 optimization at lower doping levels.

\begin{figure}[h!]
    \vspace{1em}
    \centering
    \hspace{-15pt}
    \ifdefined\Publication
    \includegraphics[width=0.46\textwidth]{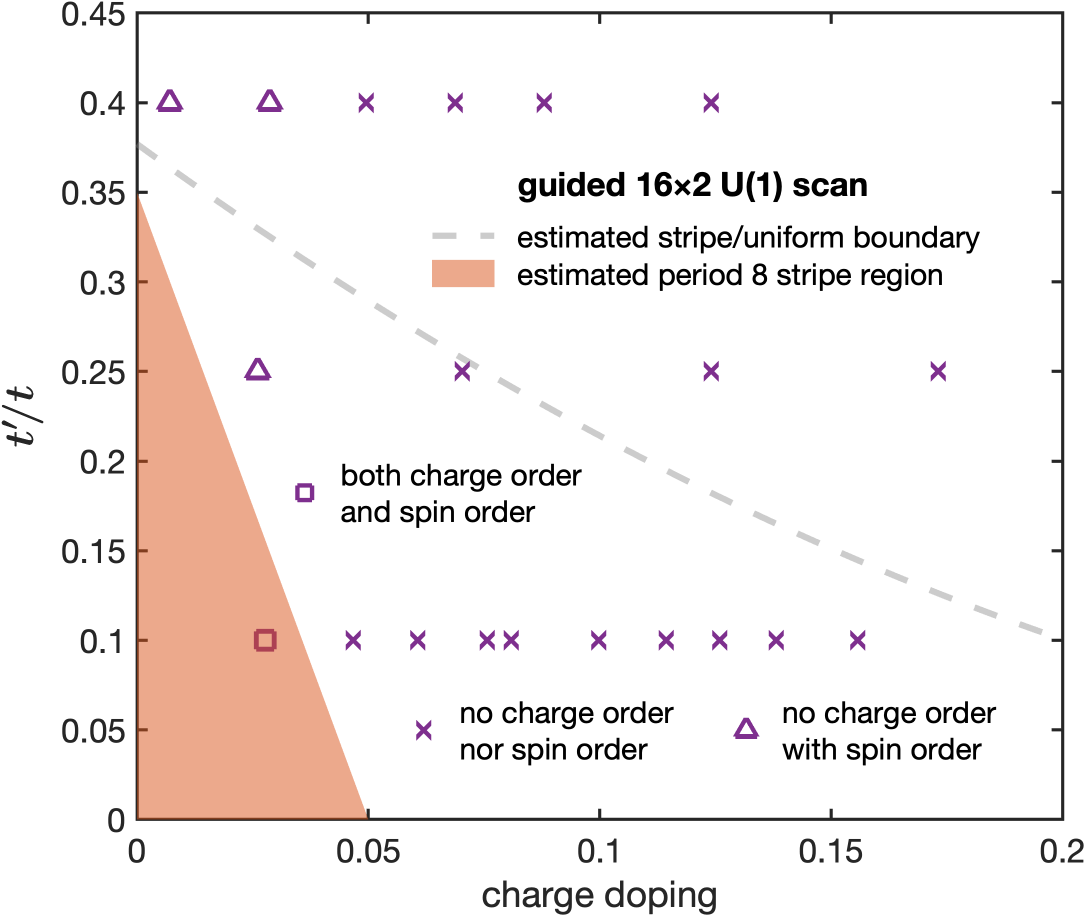}\\
    \else\includegraphics[width=0.46\textwidth]{img/LPStripe.pdf}\\ \fi
    \begin{minipage}{0.45\textwidth}
    \caption{16×2 guided $\mathrm{U}(1)$ scan at $\tnnns/\tnn = 0.1,0.25,0.4$, respectively. Cross markers represent the 16×2 $\mathrm{U}(1)$ iPEPS simulations maintaining uniform states devoid of charge and spin orders. Circle markers denote the 16×2 $\mathrm{U}(1)$ iPEPS simulations sustaining uniform in charge orders but displaying spin orders. For even lower doping, charge orders start to emerge, indicated by an orange shaded region.}
    \label{LPStripe}
    \end{minipage}
\end{figure}

The outcome of this process is illustrated in Fig.~\ref{LPStripe}. At the majority of scan points where $\delta\!>\!0.05$, the ground state derived from further 16×2 $\mathrm{U}(1)$ iPEPS optimization remains physically the same as the uniform state produced by guided 2×2 $\mathrm{U}(1)$ iPEPS (cross markers). As the doping level decreases, spin orders begin to appear (plus markers), followed by the charge orders (square marker). The pairing order is suppressed once the spin orders are fully developed. The estimated region for period 8 stripes is indicated as an orange shaded area in Fig.~\ref{LPStripe}. This region clearly resides beneath the estimated stripe/uniform boundary as derived in the main text. Consequently, this boundary can be confidently considered as a non-superconducting/superconducting demarcation for $\delta>0.05$.